\documentclass[10pt]{article} % For LaTeX2e
\usepackage[preprint]{tmlr}

\usepackage{amsmath,amsfonts,bm}

\def\eqref#1{equation~\ref{#1}}
\def\1{\bm{1}}

\DeclareMathAlphabet{\mathsfit}{\encodingdefault}{\sfdefault}{m}{sl}
\SetMathAlphabet{\mathsfit}{bold}{\encodingdefault}{\sfdefault}{bx}{n}

\usepackage{hyperref}
\usepackage{url}
\usepackage[pdftex]{graphicx}
\usepackage{arydshln} % for dashed line
\usepackage{multirow}% http://ctan.org/pkg/hhline
\usepackage{hhline}
\usepackage{ehhline}
\usepackage{dashrule}
\usepackage{amssymb} 
\usepackage{pifont} 
\usepackage{amsmath,amsfonts}

\usepackage[caption=false,font=normalsize,labelfont=sf,textfont=sf]{subfig}
\usepackage{textcomp}
\usepackage{stfloats}
\usepackage{subfig}
\usepackage{url}
\usepackage{verbatim}
\usepackage{graphicx}
\usepackage{cite}
\usepackage{microtype}
\usepackage{blindtext}
\usepackage{tikz}
\usepackage{cite}
\usepackage{graphicx}
\usepackage{subcaption}
\usepackage{xcolor}
\usepackage{url}
\usepackage{pdflscape}
\usepackage{mathtools}

\usetikzlibrary{shapes,arrows,positioning,calc,decorations.pathmorphing,decorations.markings,shapes.geometric}
\usepackage{textcomp}
\usepackage{color,soul} % highlighting etc
\usepackage{amsmath}
\usepackage{fixltx2e}
\newlength{\pagewidth}
\title{Reconciling Privacy and Explainability in High-Stakes: A Systematic Inquiry
}

\author{\name Supriya Manna \email reachsmanna@gmail.com \\
        \addr SRM University AP, India
      \AND
      \name Niladri Sett \email settniladri@gmail.com \\
      \addr SRM University AP, India
      }

\def\month{05}  % Insert correct month for camera-ready version
\def\year{2024} % Insert correct year for camera-ready version
\def\openreview{\url{https://openreview.net/forum?id=DQqdjPcE6g}} % Insert correct link to OpenReview for camera-ready version

\begin{document}

\maketitle

\begin{abstract}
Deep learning’s preponderance across scientific
domains has reshaped high-stakes decision-making, making it
essential to follow rigorous operational frameworks that include
both Right-to-Privacy (RTP) and Right-to-Explanation (RTE).
This paper examines the complexities of combining these two
requirements. For RTP, we focus on ‘Differential privacy’ (DP),
which is considered the current gold standard for privacy-
preserving machine learning due to its strong quantitative
guarantee of privacy. For RTE, we focus on post-hoc explainers:
they are the go-to option for model auditing as they operate
independently of model training. We formally investigate DP
models and various commonly-used post-hoc explainers: how
to evaluate these explainers subject to RTP, and analyze the
intrinsic interactions between DP models and these explainers.
Furthermore, our work throws light on how RTP and RTE can
be effectively combined in high-stakes applications. Our study
concludes by outlining an industrial software pipeline, with the
example of a wildly used use-case, that respects both RTP and
RTE requirements.
% The integration of deep learning into diverse high-stakes scientific applications demands a careful balance between \textbf{P}rivacy and \textbf{E}xplainability. This work explores the interplay between two essential requirements: Right-to-Privacy (RTP), enforced through differential privacy (DP)—the gold standard for privacy-preserving machine learning due to its rigorous guarantees—and Right-to-Explanation (RTE), facilitated by post-hoc explainers, the \textit{go-to} tools for model auditing. We systematically assess how DP influences the applicability of widely used explanation methods, uncovering fundamental intricacies between privacy-preserving models and explainability objectives. Furthermore, our work throws light on how RTP and RTE
% can be reconciled in high-stakes. Our study, with the example of a wildly used use-case, concludes by outlining
% a novel software pipeline that upholds RTP and RTE requirements.
\end{abstract}

\section{Introduction}
Deep learning has achieved massive success over the past decade in several domains and high stakes \citep{dong2021survey}. It has been a cornerstone ever since in almost all aspects of scientific discoveries. However, it is to be noted that deep learning, although predominant in today's scientific understandings, comes with its inherent shortcomings \citep{talaei2023deep, saeed2023explainable, boulemtafes2020review}. In this paper, we shall investigate the intricacies of two such pivotal shortcomings: \textbf{Privacy} and \textbf{Explainability}. We shall thereafter substantiate our findings with a famous use case from previous studies.

Firstly, the deep models, although achieve superior performance across domains, are inherently prone to sensitive data leakage \citep{boulemtafes2020review}. It has been extensively shown that the models can leak information about the data with which it has been trained. Even, especially in the biomedical domain, it has been shown that simple `linkage' attacks can exploit anonymized electronic health records \citep{sweeney2015only}. Due to the inherently vulnerable nature of deep models, it is not only prone to severe privacy attacks including membership inference (MIA) \citep{shokri2017membership, hu2022membership}, model stealing \citep{oliynyk2023know}, model inversion \citep{fredrikson2015model, wang2021variational, veale2018algorithms} etc but also impose a threat on the applicability of deep learning in real-world applications.

Secondly, the deep models are \textit{hard} to explain \citep{saeed2023explainable}. Neural nets are nonlinear systems, often try to learn the (approximate) distribution of the training data. However, once a model is deployed in an \textit{open-world} setting \citep{zhu2024openworldmachinelearningreview}, there is no gold label to cross-check the prediction obtained as a result, the subtle notion of \textit{trust} on the prediction is pivotal to move forward with the same. But as these models are excessively critical, how did the model come to the prediction is a non-trivial phenomenon \citep{jentzen2023mathematicalintroductiondeeplearning}.

To tackle the first problem, researchers have developed several strategies for privacy-preserving machine learning \citep{boulemtafes2020review}. This includes homomorphic encryption \citep{pulido2020survey}, PATE \citep{papernot2018scalableprivatelearningpate}, differential privacy (DP) \citep{Abadi_2016} etc; depending on the setting and motivation we incorporate these methods to make the model \textit{robust} against diverse privacy attacks. However, differential privacy (DP), over the years, has been established as a \textit{gold-standard} for privacy-preserving machine learning \citep{Blanco_Justicia_2022, suriyakumar2021chasing} mainly due to its \textit{worst case guarantee} against \textit{any} inference attacks \citep{Abadi_2016, dwork2014algorithmic}. Furthermore, other types of privacy attacks are not as mature and/or successful as MIA is \citep{Rigaki_2023}. In this paper, we have specifically worked on DP models\footnote{Throughout the paper we use DP model(s)/explanation(s) and private model(s)/explanation(s) interchangeably.} with diverse \textit{privacy guarantee}s  aka $\epsilon$ \citep{dwork2014algorithmic}.

On the other hand, to tackle the explainability problem, researchers have developed an array of methods to audit the model decisions. This includes (but not limited to) local and global explainability methods, inherently interpretable machines, etc \citep{saeed2023explainable}. Among these, local post-hoc explainers have emerged as popular and widely adopted tools for model auditing in recent times due to their `plug-and-play' nature \citep{bhatt2020explainablemachinelearningdeployment}. In this work, we employ five popular post-hoc explainers relevant to our use case to obtain explanations.

Although privacy and explainability aspects have been vastly explored independently, there is little to no work incorporating them together. Despite the rapid advancements in AI, integrating privacy and explainability in high-stakes domains remains an unsolved and pressing challenge. Existing research has treated privacy and explainability as separate challenges, leaving a significant gap in understanding their interplay. This gap becomes critical in high-stakes applications where both are non-negotiable. Researchers in privacy-preserving machine learning have investigated several aspects of differential privacy such as the scalability of DP models \citep{beltran2024towards}, their fairness \citep{fioretto2022differential}, robustness \citep{tursynbek2020robustness},  privacy-utility trade-off \citep{zheng2024rethinking} etc. On the other side, researchers have worked on different aspects of explainable AI (XAI) such as faithfulness \citep{lyu2024towards}, robustness \citep{alvarezmelis2018robustnessinterpretabilitymethods}, and the quality of explanations \citep{zhou2021evaluating} to name a few. In this paper, we take the first comprehensive step toward bridging the divide between privacy-preserving models and explainable AI in high-stakes domains. We explore the unique challenges that emerge when attempting to integrate privacy and explainability aspects in high-stakes applications. We investigate the underlying causes of these challenges, examine the trade-offs involved, and discuss key considerations necessary for developing frameworks to successfully incorporate these two critical aspects effectively.

Integrating privacy and explainability is non-trivial, but addressing these issues is crucial for ensuring trustworthy and effective AI systems in high-stakes. In this context, we argue that it is equally important to identify use cases where achieving `the best of both worlds' is not just desirable but a \textit{requirement}. For example, if the training data is open source or an interpretable model (e.g. decision trees) is considered for the study, our research may not be well-motivated or the findings are not `worthy'. This is why, for this paper, we have considered the well-established use-case of disease detection from chest x-ray \citep{al2024chest}. The reason to choose the same is that medical records are (almost) always subject to privacy preservation and it is of paramount importance that, with explanations, the stakeholders including both physician and patients can \textit{trust} the model predictions. Post-hoc explainers, which we use in this paper, have been extensively utilized for our chosen and similar use cases in past \citep{e2024evaluating,saxena2022artificial,ifty2024explainable}.

The dual mandates of the Right-to-Privacy (RTP) \citep{thomson1975right} and the Right-to-Explanation (RTE) \citep{vredenburgh2022right}, increasingly imposed by governments for high-stakes AI applications, underscore the urgency of addressing these intertwined imperatives. Despite their critical importance, prior research lacks a comprehensive and systematic feasibility study that explores the practical integration of these rights within AI systems. Fundamental questions remain unanswered: How should we evaluate the quality of explanations for DP models? Can existing popular local post-hoc explainers reliably function in this setting, or do they falter under the constraints of differential privacy? If they fail, what alternative methods can generate comparable private explanations? Our work makes a fundamental contribution by systematically addressing these pressing questions, providing not only foundational clarity but also actionable insights grounded in rigorous empirical analysis. Formally, our central research question in this paper is:

\begin{center}
    $\blacksquare$ \underline{Do DP models and post-hoc explainers \textit{go together}}?
\end{center}

Our contributions are as follows:
\begin{itemize}

 \item We propose the desiderata for private explanations to follow;
 \item We investigate the interplay between the DP models and post-hoc explainers, subject to the desiderata proposed;
  \item We conduct extensive experiments and  report our findings: we consider three types of widely used CNN models, six distinct $\epsilon$ values to train them on, and five popular post-hoc methods for our experiment;
 \item We present a rigorous study on the mechanistic interpretation of the DP models; depicting the interplay of the post-hoc explainers with DP models;
 \item We propose a novel framework that can achieve RTP and RTE together for high-stakes applications; we exemplify the same by outlining an end-to-end privacy-preserving pipeline for our use case.

 \end{itemize}

% \begin{enumerate}
%  \item\textbf{Localization Assumption}
%  \item\textbf{Feasibility Test for Using Post-hoc Explainers with DP Models}
%  \item\textbf{The Alternative Setup}
% \end{enumerate}

The rest of the paper is organized as follows. We brief the preliminary concepts of DP and post-hoc explainers in Section \ref{pril}. We point out the potential privacy breaches, followed by an extensive discussion on several quantifiable notions for explanations' quality,  and their inherent shortcomings in Section \ref{TPAB}. We formally introduce our proposed postulate and quantifiable measures for the same in Section \ref{LA-PIS}. We detail our experimental setup in Section \ref{ESET}. We discuss our findings in Section \ref{DIS}. We mechanistically interpret the models to explain our findings in Section \ref{EXP}. Next, we explore Local Differential Privacy and its applicability in our context, in Section \ref{LDP-1}. We outline our novel software pipeline to \textit{reconcile} privacy and explainability in Section \ref{swe}. We present the related works in Section \ref{RELW}. Finally, we conclude in Section~\ref{CONC}.

\section{Preliminaries}\label{pril}
\textbf{Privacy-preserving machine learning} is the building block for RTP in modern day ML systems. Researchers have developed numerous strategies for privacy-preserving machine learning, including homomorphic encryption, PATE, and differential privacy (DP) \citep{boulemtafes2020review, papernot2018scalableprivatelearningpate, Abadi_2016}. Among these, DP has emerged as the \textit{gold standard} due to its strong \textit{worst-case guarantee} against inference attacks \citep{Blanco_Justicia_2022, dwork2014algorithmic}. In this work, we focus on DP models with varying privacy guarantees.

A randomized mechanism $\mathcal{F}$ satisfies $(\epsilon, \delta)$-DP if, for any two neighboring datasets $D$ and $D'$ differing in at most one record, and for all measurable subsets $\mathcal{S}$ of the output space,
\[
\Pr[\mathcal{F}(D) \in \mathcal{S}] \leq e^\epsilon \Pr[\mathcal{F}(D') \in \mathcal{S}] + \delta.
\]
Here, $\delta$ represents the probability of the mechanism failing to provide privacy guarantees. Specifically, it accounts for the small chance that the added noise does not sufficiently obscure the presence or absence of an individual in the dataset. Smaller values of $\delta$ are preferred.

DP comes with a few interesting properties such as sequential composition, parallel composition and, post-processing. In our study, post-processing is most relevant which lets the user perform arbitrary operations on the output of a DP mechanism without hampering its privacy \citep{dwork2014algorithmic}.

To make a model differentially private (DP), it undergoes privacy-preserving training, with DP-SGD being the most prominent method \citep{Ponomareva_2023}. It introduces Gradient clipping ensuring bounded contributions of individual data points and noise to the gradients during each update to ensure differential privacy \citep{Abadi_2016}. Gaussian noise is commonly used due to its well-analyzed privacy guarantees. Alternatively, laplacian noise is employed in some contexts where bounded sensitivity is easier to calculate \citep{dwork2014algorithmic}. For DP-SGD, gaussian noise is predominantly used whereas, Laplacian noise is popular in Local Differential Privacy (LDP) \citep{dwork2014algorithmic,Ponomareva_2023}.

Local Differential Privacy (LDP) provides privacy guarantees at the data source. A mechanism $\mathcal{F}$ satisfies $\epsilon$-LDP if, for any two inputs $x$ and $x'$ and all outputs $y$,
\[
\Pr[\mathcal{F}(x) = y] \leq e^\epsilon \Pr[\mathcal{F}(x') = y].
\]

In this work, we denote the non-private model as $\mathcal{M}$. We obtain its private counterpart $\mathcal{M'}$ by retraining with DP-SGD.

\textbf{Explainable machine learning} is crucial for Right-to-Explainability (RTE). As mentioned earlier, this paper focuses on local, post-hoc explainers (denoted as $\mathcal{I}$). Broadly, these explainers can be classified into two categories: Perturbation-based methods and Gradient-based methods. Both types of methods take a model ($g$) and a datapoint ($x$) as input and output a feature attribution score(s) against $g(x)$. A feature attribution score (FAS) (or a feature attribution vector) consists of scores assigned to individual features of the given input representing their importance towards the classification the given model comes up with. A positive attribution score implies a feature has positively contributed towards the classification and a negative attribution score shows that a feature does not positively contribute towards the classification. Given an explainer and a datapoint ($x$), we denote the FAS which explains the prediction of $\mathcal{M}$ and $\mathcal{M'}$ as \textit{s} and $s'$ respectively; given $\mathcal{M}(x) =$ $\mathcal{M'}(x)$.

In our use case, \textit{s} (and $s'$) are typically computed on a per-pixel or per-element basis, and they generally match the dimensions of the input or the corresponding intermediate layer. In the next section, we shall discuss the non-private setup and potential security breaches involved.

\section{Two Side of the Coin: Privacy and Explainability}
\label{TPAB}
\subsection{The Privacy Aspect}\label{TPA}
% There is an array of privacy mechanism such as Differential Privacy (DP), Private Aggregation of Teacher Ensembles (PATE), Local Differential Privacy (LDP), Homomorphic Encryption (HE), Secure Multi-Party Computation (SMPC), etc. Unlike HE and SMPC, which provide cryptographic security without quantitative privacy budgets, DP, PATE, and LDP offer formal privacy guarantees via a privacy budget. PATE consists of ensemble of disjoint `teacher' model for predict the outcome of the given input. Firstly, it introduces huge computational overhead. Secondly, due to the inherent ensemble model of the setting, the final explanation of the given input is not consensus and potentially \textit{socially misaligned} \citep{jacovi2021aligning}. Thirdly, due to the disjoint nature of the models, it is excruciatingly hard for the developers to upgrade the software and security patches. In this paper, we have focused on the notion ofDPthroughout; it has also been a recent point of interest for the bio-informatics community. A detailed discussion on the privacy measures and their applicability are presented in Appendix A. In the next section, we will try to understand different kinds of security breaches.
The non-private setup consists of an adversary $\mathbf{A}$, having access to a data-point $x$, a non-private model $\mathcal{M}$, the output vector $\textbf{V}$ obtained from $\mathcal{M}$ for $x$, an explainer $\mathcal{I}$ generating a feature attribution score \textit{s} against $\mathcal{M}(x)$.

From the point of $\mathbf{A}$, we identify two distinct ways of attacking and leaking the information from the trained model\footnote{As in this paper we are working with DP models, we have considered MIA as the only type of potential privacy attack for simplicity. However, it has been shown that other classes of privacy attacks (model stealing, model inversion, etc) are also very much possible with the two ways of attacking mentioned, and DP potentially can safeguard against a few other attacks besides MIA \citep{Rigaki_2023}. However, these fall beyond the scope of the present study.}:
\begin{enumerate}
    \item For a given data-point $x$, assuming $\mathbf{A}$ has access to the output vector $\textbf{V}$ from the trained model $\mathcal{M}$, $\mathbf{A}$ can perform an MIA on the model \citep{hu2022membership}.
    \item Assuming $\mathbf{A}$ has access to the feature attribution score \textit{s} of $x$ generated by $\mathcal{I}$ subject to $\mathcal{M}(x)$, $\mathbf{A}$ can leverage \textit{s} to execute an MIA on $\mathcal{M}$ \citep{shokri2021privacy}, especially when $\mathcal{I}$ is \textit{faithful}.
\end{enumerate}

However, as $\mathcal{M'}$ is the DP counterpart of $\mathcal{M}$, it inherits the post-processing property \citep{dwork2014algorithmic} which makes \textit{any} mechanism including obtaining \textbf{V}, $s'$, from $\mathcal{M'}$ to produce results which are also DP, \textit{i.e.}, \textit{any} post-hoc explanations for an ($\epsilon, \delta$)-DP model also satisfy ($\epsilon, \delta$)-DP for the training dataset \citep{patel2022model}. Thus, using $\mathcal{M'}$ we can mitigate both of the breaches.

\subsection{The Explainability Aspect}\label{TEA}
As discussed earlier, in this study, we are exclusively considering local post-hoc explainers \citep{lundberg2020local,huber2021local, lundberg2017unified}. The \textit{quality} these local explanations are judged across several parameters \citep{hedstrom2023quantus}: \texttt{Faithfulness} \citep{lyu2024towards}, \texttt{Robustness} \citep{alvarezmelis2018robustnessinterpretabilitymethods}, \texttt{Localization} \citep{zhang2018top,theiner2022interpretable}, \texttt{Complexity} \citep{chalasani2020concise,nguyen2020quantitative}, \texttt{Randomization} \citep{adebayo2018sanity}, are mostly accentuated in previous studies.

Unequivocally, \texttt{faithfulness} is the most crucial among all \citep{lyu2024towards}. \texttt{Faithfulness} is loosely defined as how well the explainer reflects the underlying reasoning of the model and has been extensively quantified in diverse ways \citep{li2023mathcal,hedstrom2023quantus}: \texttt{Sufficiency (SF)} \citep{dasgupta2022framework}, \texttt{Infidelity (IF)} \citep{yeh2019fidelity}, \texttt{Insertion/Detection AUC (I/D-AUC)} \citep{petsiuk2018rise}, \texttt{Pixel Flipping (PF)} \citep{bach2015pixel}, \texttt{Iterative Removal of Features (IROF)} \citep{rieger2020irof}, \texttt{Ordered Perturbation} based metrics \texttt{(OPs)} \citep{samek2016evaluating} are to mention a few. However, the evaluation metric(s) are not necessarily flawless, which we're going to discuss next.
\subsection{Pitfalls of the Evaluation Metrics}
Firstly, in the perturbation-based metrics such as \texttt{PF}, \texttt{I/D-AUC}, \texttt{IROF}, \texttt{IF} etc while applying operations such as flipping the pixels, inserting and/or deleting features, performing \texttt{meaningful perturbation} on the input space to generate synthetic inputs as a part of their evaluation process, do not crosscheck whether the generated input is out-of-distribution with respect to the trained model \citep{hase2021out, chang2018explaining}. Secondly, previous studies have extensively shown that test-time input ablation is often prone not only to generating out-of-distribution (OOD) synthetic input \citep{hase2021out,haug2021baselines} but also are \textit{socially misaligned} \citep{jacovi2021aligning}. Thus, metrics for example \texttt{I/D-AUC}, \texttt{IROF} etc are suspected to be severely misleading; even metrics such as \texttt{SF} are also substantially constrained and do not provide a universal overview of \texttt{faithfulness}. Thirdly, metrics such as \texttt{OPs} are often excessively similar to the mechanics of the explainer itself. Instead of evaluating \texttt{faithfulness}, these methods primarily compute the similarity between the evaluation metric and explanation techniques, assuming the evaluation metric itself to be the ground truth \citep{ju2021logic}. Li et al. \citep{li2023mathcal} has acknowledged the same in their benchmark $\mathcal{M}^4$ that LIME \citep{ribeiro2016should} and \texttt{OPs} are methodologically similar thus, the evaluation maybe skewed. Lastly, all these evaluations are based on naive assumptions (e.g.: erasure \citep{jacovi2020towards}), derived from a set of seemingly valid observations. Therefore, these quantitative metrics are not necessarily axiomatically valid. Even axiomatic necessary tests such as \texttt{Model Parameter Randomization test} \citep{adebayo2018sanity} happen to have a set of empirical confounders \citep{yona2021revisiting,kokhlikyan2021investigating,bora2024why}. To the best of our knowledge, there has not been any universally valid necessary and sufficient approach for faithfulness evaluation \citep{lyu2024towards}.

Furthermore, the absence of ground truth for models' reasoning makes it an open challenge to universally quantify faithfulness and evaluate a faithfulness measure precisely, i.e. when a universal ground truth for reasoning is not available, axiomatically quantifying how much a faithfulness measure is more reliable than others is challenging; however, in the existing literature, faithfulness measures evolve by addressing the discovered shortcomings of their predecessors \citep{lyu2024towards}. This also leads to the well-known disagreement problem \citep{krishna2022disagreementproblemexplainablemachine}: even if explainers are \texttt{faithful}, their inherent mechanisms for calculating feature importance can lead to different explanations. In the subsequent section(s), we'll try to figure out whether we can bypass these limitations for our study.
%For instance, Han et al. \citep{han2022explanation} demonstrated that while a subset of commonly used XAI methods are all local function approximators, they employ different kernel and loss functions. Consequently, like \texttt{faithfulness}, other parameters for evaluating explainers are also contentious. For example, \texttt{adversarial robustness} (of the explainer) assumes that similar inputs with similar outputs should yield similar explanations \citep{sinha2021perturbing,ivankay2022fooling}. However, Ju et al. \citep{ju2021logic} has empirically shown that the change in attribution scores may be because the model’s reasoning process has genuinely changed, rather than because the attribution method is unreliable. Moreover, this assumption is mainly valid when the model is \textit{astute} \citep{bhattacharjee2020non} and doesn’t necessarily apply to explainers that don’t perform local function approximation for feature importance estimation \citep{han2022explanation}. 

\subsection{Aspiration for the Alternatives}
Acknowledging these inherent limitations in current evaluation methods, we propose two key remarks:
\begin{enumerate}
\item \textbf{Remark 1.} Expert oversight should determine whether an explanation is \textit{suitable} for high-stakes applications.
\item \textbf{Remark 2.} Explanations should align with local constraints and contexts, even when (so-called) \textit{faithfulness} cannot be measured reliably.
%\footnote{For example in our case, local constraint is the similary between of the private and vanilla explanation, local context is that the private explanation can be used as a \textit{proxy} to its vanilla counterpart, subject to physicians' approval.}
\end{enumerate}
To address the first requirement, in our use case we ensure that concerned physicians first receive X-ray images along with predictions and explanations. Results are only communicated to patients after the physician has completed a formal review and certified both the prediction and the explanation.
% Human-machine collaboration is a cornerstone of Industry 5.0 \citep{xu2021industry,leng2022industry}; thus, this protocol not only preserves this norm but also helps to collect feedback from doctors, particularly with respect to uncertified samples on a periodic basis. This feedback is instrumental in software upgradation and better alignment with the needs and data specific to different geographical regions.\newline

For the second remark, we are introducing the \texttt{localization assumption} (LA) and quantifying the same with a class of measures we collectively named the Privacy Invariance Score (PIS) for explanations.
\section{Localization Assumption \& Privacy Invariance Score}\label{LA-PIS}
% Privacy in machine learning has become paramount, with policies such as \_ and \_ emphasizing the use of privately trained models in high-stakes applications, including biomedical software \citep{}. Differential privacy \citep{} offers quantitative privacy guarantees, but it is well established that private models often exhibit reduced accuracy compared to their non-private counterparts \citep{}. As noise is induced in the gradient during training, private and non-private models can differ substantially, depending on the allocated privacy budget.

Right-to-Privacy (RTP) \citep{thomson1975right} and Right-to-Explanation (RTE) \citep{vredenburgh2022right} are two inalienable aspects of modern-day ML software. We have already mentioned that previous works have shown potential security breaches leveraging the explanation in section \ref{TPA}. So, RTE can hamper RTP but is the converse true? If yes, how can we quantify the same? We first propose the desiderata for explanations in this setting and then run extensive experiments to quantitatively \textit{judge} the explainers.

\subsection{Description of the setting}
It is shown that adversaries can leverage \textit{sensitive} information from explanations but due to post-processing of DP, any DP model will always output a private explanation \citep{dwork2014algorithmic}. We want to check the extent to which the explanations from a DP model can be used as a proxy for that of the non-private model here. Formally, for a non-private model $\mathcal{M}$, its private counterpart $\mathcal{M'}$; consider a set of post-hoc explainers $\mathcal{E}$ used for auditing $\mathcal{M}$ and $\mathcal{M'}$. For an input $x \in X$ (where $X$ is the valid input space), each $\mathcal{I} \in \mathcal{E}$ produces explanations $s$ and $s'$ of $x$ for $\mathcal{M}(x)$ and $\mathcal{M'}(x)$ respectively, Privacy Invariance Score \textbf{(PIS)} over a given tuple $(\mathcal{M}, \mathcal{M'}, x, \mathcal{I})$ is defined as follows.

\textbf{Definition 1.} Privacy Invariance Score \textbf{(PIS)}: Given a tuple $(\mathcal{M}, \mathcal{M'}, x, \mathcal{I})$, given $\mathcal{M}(x)$ = $\mathcal{M'}(x)$, PIS is defined as $sim(s, s')$, where $sim(\cdot, \cdot)$ is a \textit{similarity} measure, and $s = \mathcal{I}(\mathcal{M}, x)$ and $s' = \mathcal{I}(\mathcal{M'}, x)$.

\subsection{Localization Assumption}
\label{Sec-LA}
As discussed earlier, different works evaluated explainers in various ways. These ad-hoc norms are often unique to each paper and inconsistent \citep{jacovi2020towards}. However, practitioners have proposed some necessary axiomatic desiderata \citep{lyu2024towards} such as the well-established \texttt{Implementation Invariance (II)} criterion proposed by Sundararajan et al.~\citep{sundararajan2017axiomatic}.

According to \texttt{II} if $\mathcal{I}$ is \textit{faithful} and $\mathcal{M}(x) = \mathcal{M'}(x)$ $\forall x \in X$, i.e. $\mathcal{M, M'}$ are \textit{functionally equivalent} (FE) then $s = s'$ $\forall x$.
Commonly used explainers like \texttt{Integrated Gradient} \citep{sundararajan2017axiomatic}, \texttt{SmoothGrad} \citep{smilkov2017smoothgrad}, \texttt{DeepLift} \citep{li2021deep}, \texttt{layerwise relevance propagation (LRP)} \citep{montavon2019layer} etc had been extensively accessed based on \texttt{II}~\citep{sundararajan2017axiomatic}.

Following these lines of prior explainability research, Jacovi et al. later formally proposed \texttt{The Model Assumption} (\texttt{MA}) \citep{jacovi2020towards}:

\textit{``Two models will make the same predictions if and only if they use the same reasoning process.''}

Nevertheless, in our setting, as the accuracy of $\mathcal{M'}$ is generally less \citep{Abadi_2016}, and we cannot say $\mathcal{M}$ and $\mathcal{M'}$ are FE \citep{sundararajan2017axiomatic}. Furthermore, as $X$ in practice can be arbitrarily large (theoretically could be countably infinite), quantifying whether an explainer ($\mathcal{I}$) is \textit{sufficiently} trustworthy or not is often impractical. Also, finding a counterexample that violates the condition implying $\mathcal{I}$ is not faithful is computationally expensive. Hence, we modify \texttt{MA} and propose the \texttt{localization assumption}  for evaluating the explainers' quality in our setting.

First of all, in both the \texttt{II} and \texttt{MA} we do not advocate comparing any arbitrary models trained on the same data, having the same $X$ to compare. For example, for a finite $X$, a trained neural network and a decision tree could have the same prediction $\forall x \in X$. That doesn't mean their \textit{reasoning} is similar as the algorithms themselves are different. However, the private model $\mathcal{M'}$, in our setting, has identical architecture to that of $\mathcal{M}$. The fundamental goal of differential privacy is assumed to be masking individual contributions of the training set rather than completely changing its overall reasoning. However, as noise is induced in the gradient during the training, the parameters are expected not to remain entirely the same in the private model. Therefore, we assume that their reasoning should primarily be \textit{similar}, if not the \textit{same}. Formally, our adapted version of the MA is:

\textbf{Assumption 1 The \texttt{Localization Assumption} (LA).} \textit{For a given tuple $(\mathcal{M}, \mathcal{M'}, x, \mathcal{I})$ having $\mathcal{M}(x) = \mathcal{M'}(x)$, $sim(s, s') >= \theta$. Where $\theta$ is a predefined similarity threshold.}\footnote{It is worth nothing that LA and \texttt{Localization} discussed in \ref{TEA} are different.}

In this context, since we are using a proxy model (in our case, a DP model) as a substitute for the original model (the non-private model), we argue that beyond satisfying LA, the system must also meet two additional requirements:
\begin{enumerate}
    \item \textbf{Performance comparability} (\textit{Perf Comp.}): The proxy model should achieve comparable performance metrics to the original model. For this study, we focused on accuracy (acc) since in our test set, we put equal weightage on all classes and there is no class imbalance (further details are provided in Section \ref{DIS}). Specifically, we report $Acc_{\mathcal{M'}/\mathcal{M}}=\frac{\text{acc. of }\mathcal{M'}}{\text{acc. of }\mathcal{M}}$ for our experiments.
    \item \textbf{Alignment with the original model} (\textit{Alignment}): The proxy model must closely align with the original model in its predictions. In our study, we measure the same with the agreement on the ``hard predictions" obtained from $\mathcal{M}$ and $\mathcal{M'}$ over the test set. (Dis)agreement has been extensively used in previous studies due to its its simplicity and interpretability \citep{klabunde2024similarityneuralnetworkmodels}.
\end{enumerate}
These two requirements are critical for ensuring the effectiveness of the proxy model. \textit{Perf Comp.} is necessary because a significant drop in the proxy model's performance would undermine its utility as a stand-in for the original model, especially in high-stakes applications. \textit{Alignment} is equally important because, without sufficient agreement on predictions, the proxy model will fail to replicate the original model's decision-making patterns, making it ineffective as a proxy. We refer to a proxy model that meets both these requirements as \textit{functionally comparable} w.r.t. the original model.

%add why and how Asssumption 1 is valid for all such cases where we expect to replace a model with its goal oriented proxy model (such as adversarialy robust, DP). We argue that functional similarity is a crude estimate in such cases.
Next, we investigate how we can measure the `similarity' described in the \texttt{Localization Assumption}.

\subsection{The Notion of Similarity}\label{SIM}
As previously discussed in PIS, we aim to measure the similarity between pairs of explanations; it is essential to account for two key factors here: the context of comparison and human understanding of that comparison.
% Formally, let $\mathbf{l}$ represent an explanation vector from explainer $\mathbf{E}$ for an input $\mathbf{x}$ and a non-private model $\mathbf{f}$, while $\mathbf{l'}$ denotes the corresponding private counterpart from $\mathbf{f'}$.
In this framework, given $(\mathcal{M}, \mathcal{M'}, x, \mathcal{I})$ and $\mathcal{M}(x) = \mathcal{M'}(x)$, we seek to investigate two primary aspects:

\begin{itemize} \item To what extent do \textit{s} and $s'$ \textit{agree}? \item Wherever they agree, what is the degree of that agreement? \end{itemize}
Firstly, explanations typically include both positive and negative attributions\footnote{\texttt{Saliency}, for example, outputs only positive attributions as absolute values.}, where positive attributions favor the classification, and negative ones do not. To evaluate explanations' similarity, we first compute a disagreement score (DS, measured in \%), by measuring the mismatch of corresponding attribution type (negative or positive) in ($s, s'$). DS is a (crude) measure for assessing how much the reasoning of the models, as indicated by $\mathcal{I}$, diverges at the pixel or element level. In several highstakes (e.g. biomedical), as both position and attribution are important, we employ DS primarily as a sanity check in our case study with a threshold (hyperparameter) of 15\%.

%---------------------------------------I concate a shorter version of the below with the paragraph above---------------------------------------------------
%From a human evaluation perspective though, analysis on DS is context-sensitive. While evaluating DS may not be crucial in all applications, it becomes critically important in high-stakes domains, such as biomedical software, where discrepancies could lead to significant confusion. In these contexts, both position and attribution of information are vital for accurate decisions. To address this, our system evaluates disagreement at an atomic level, ensuring precision and thoroughly scrutinizing the explainer when a specific threshold of disagreement is identified for an image. In our study, we set the DS threshold 15\%. We employ DS primarily as a sanity check in our experiment for the comparability of model explanations.
Secondly, after measuring DS and eliminating the explainers not adhering the threshold, we assess the attribution correlation for the rest. We start by segregating the positive attributions from the negative ones, as the former contribute positively to the classification, while the later may (partially or completely) contribute to other classes or contribute negatively to the output class or a combination of these two \citep{ancona2017towards,samek2017explainableartificialintelligenceunderstanding}. Consequently, we compute the Kendall's tau between pairwise positive pixels of \textit{s} and $s'$ to measure PIS. While alternative notions for similarity like cosine similarity or \( L_{P} \) norms exist, cosine is unsuitable in high dimensions due to near-orthogonality of random vectors \citep{MOHAMMADI2022103515,6771362}, and \( L_{P} \) norms require normalization, losing one degree of freedom unlike correlation.

\section{Experimental Setup}\label{ESET}

\subsection{Description of the Dataset}

Our dataset comprises 2,000 Pneumonia cases sourced from the Chest X-ray dataset by \citep{prashant2682020chestxray}, and 2,000 TB cases randomly sampled from the NIAID TB Portal Program dataset \citep{NIAID_TB_Portal}. To create the `Normal' subset, we include an equal split of 1,000 unaffected Pneumonia samples from the unaffected class in the aforementioned Chest X-ray dataset \citep{prashant2682020chestxray} and 1,000 unaffected TB samples from Rahman et al. \citep{rahman2020reliable}, totaling 2,000 normal cases. For evaluation, the test set contains 200 images for each class.

This paper, unlike prior studies, is not focused on the empirical studies of private models in healthcare \citep{naresh2023privacy,khalid2023privacy}. Instead, we concentrate on the applicability of commonly used explainers in a privacy-preserving environment. Our aim is to investigate whether RTE and RTP can be achieved simultaneously. As a result, rather than a large dataset where extensive experimentation on several types of CNNs with DP under a closely controlled environment is severely challenging and computationally demanding \citep{Ponomareva_2023}, we focus on first making a dataset of an appropriate size where we have considered the commonly available and utilized disease class from the previous studies. However, after our primary experiment, we, within the limits of feasibility, experimented with another benchmark dataset: CIFAR-10, and we arrived at a similar conclusion. Details can be found in Appendix A. %\ref{App A}.

\subsection{Choosing the Privacy Budget ($\epsilon$)}
Selecting $\epsilon$ is critical to balancing privacy requirements and utility. The theoretical school advocates for small $\epsilon$s (e.g. $\epsilon \leq 1$), offering strong worst-case guarantees (for e.g. $\epsilon = 0.1$ \textit{theoretically} limits MIA success rates to 52.5\%). In contrast, industrial settings often adopt large $\epsilon$ (e.g. $\epsilon > 7$), which, while practical, \textit{theoretically} yield MIA success rate exceeding 99\%.

This dichotomy arises from differing assumptions \citep{lowy2024does}: the theoretical school presumes adversaries with \textbf{near-complete knowledge} of training data and \textbf{uniform privacy guarantees} for all datapoints. However, recently Lowy et al. \citep{lowy2024does} highlights the impracticality of these assumptions and showed that even $\epsilon \geq 7$ can provide adequate defense against existing MIAs. 
%Furthermore, in high-stakes domains like healthcare, data is often either fully public or entirely private, with breaches leading to strict system restrictions, as seen in Italy’s 2023 temporary ban on ChatGPT due to privacy concerns\footnote{\url{https://bbc.com/news/technology-65139406}}.

While this study does not propose guidelines for choosing $\epsilon$ in biomedical applications, we adopt Google’s three-tier framework \citep{Ponomareva_2023}:\\ \textbf{Tier 1}: $\epsilon \leq 1$: strong theoretical guarantees, low utility;\\
\textbf{Tier 2}: $\epsilon \leq 10$: practical trade-offs, google advocates this tier primarily;\\
\textbf{Tier 3}: $\epsilon > 10$: high vulnerability, thus excluded from this study.

Our analysis examines $\epsilon \in \{0.4, 0.7, 1, 4, 7, 10\}$, spanning tier 1 and tier 2, to comprehensively evaluate privacy-utility trade-offs.

\subsection{Description of the Explainers}\label{xai}
We primarily selected gradient-based explainers, which are predominantly used in computer vision tasks \citep{buhrmester2021analysis} — specifically, \texttt{Saliency, SmoothGrad, Integrated Gradients, Grad-Shap} and \texttt{Grad-CAM} (details are in Appendix B). These gradient-based explainers rely on the sensitivity the model shows to the features subject to the output it obtains. In other words, these explainers leverage the gradient of the output (or any selected class of interest) w.r.t. the input features\footnote{it can either directly be the element(s) of the given input or the intermediate ones coming from a layer inside the model.}. For example, \texttt{Saliency} computes the partial derivatives of the output w.r.t. the input given, \texttt{Integrated Gradients} computes the average gradient while the input varies along a linear path from a baseline. \texttt{SmoothGrad} involves adding noise to the input image and generating multiple saliency maps, each corresponding to a slightly different noisy version of the input image; the saliency maps are then averaged to produce a `smoothed' saliency map. \texttt{Grad-CAM} also involves taking gradients of the target output with respect to the given layer. From perturbation-based explainers, we included only \texttt{SHAP} \citep{lundberg2017unified}, specifically \texttt{Grad-Shap}. \texttt{Grad-Shap}, in a sense, can be viewed as an approximation of \texttt{Integrated Gradients} by computing the expectations of gradients for different baselines. Core perturbation methods are less widely adopted for computer vision, and they tend to disobey the ``data manifold hypothesis'' \citep{fefferman2016testing}, as described in the next paragraph. Furthermore, perturbation-based methods like \texttt{LIME} \citep{ribeiro2016should} assign weights to super-pixels rather than individual pixels. As a result, we cannot directly compare the same with other methods we chose, which output per-pixel or per-element attribution values,
\texttt{Occlusion} can lead to OOD artifacts and is not widely used \citep{chang2018explaining}. Additionally, some other explainers, such as \texttt{DeepLIFT, LRP} disregard even basic faithfulness tests \citep{sundararajan2017axiomatic}.

We did not consider model-agnostic perturbation-based explainers (e.g., KernelSHAP, LIME, Occlusion) due to their relatively limited adaptation in computer vision tasks and several criticisms. Firstly, these are known to disobey the ``data manifold hypothesis'' \citep{fefferman2016testing}: Kumar et al. \citep{kumar2020problems} and Sundararajan et al. \citep{sundararajan2020many} showed these methods often operate outside the data's true distribution.  Furthermore, Shokri et al. \citep{shokri2021privacy} acknowledged the fact that “\textit{little is known on how models generally perform outside of the data manifold. In fact, it is not even clear how one would measure performance of a model on points outside of the training data distribution: they do not have any natural labels…} ”. To the best of our knowledge, not only vanilla models but also how their DP counterparts behave outside the data manifold is unknown. In fact, how different the manifold of DP models is, compared to its non-DP counterpart, and how these perturbation-based explainers may behave subject to DP models’ decision boundaries and manifolds, has not been extensively studied prior. Therefore, whether, in an ad-hoc manner, we can even employ these explainers to DP models or the suitability of these explainers in the context of DP models is not clear to us, and we’d like to leave this as an open research question to the community. However, as mentioned earlier, we considered \texttt{Grad-Shap} as it closely resembles to \texttt{Integrated Gradients}.

\subsection{Description of the Networks}
Disease detection using chest X-ray is a well-regarded case study. In previous works \citep{al2024chest}, the choice of CNNs is significantly prominent, and in this work, we have also chosen the famous CNNs that have been used in previous research. Particularly, We selected ResNet-34 \citep{he2016deep}, EfficientNet-v2 (`small' version) \citep{tan2021efficientnetv2}, and DenseNet-121 \citep{huang2017densely} for our analysis due to their competitive performance. Importantly, we ensured that the DP counterparts of these models were \textit{functionally comparable} across a wide range of values for $\epsilon$, to the best of our ability.
However, previous studies have shown that DP doesn't make the model learn all classes in a comparable fashion \citep{suriyakumar2021chasing, fioretto2022differential} as a result, we report $Acc_{\mathcal{M'}/\mathcal{M}}$ as a measure of \textit{Perf Comp.} and agreement on ``hard prediction'' to measure \textit{Alignment} on a class-wise basis, allowing for a more nuanced comparison. We detailed our experimental setup in Appendix C.

\begin{table}[h!]
\centering
\begin{tabular}{c|c|c|c|c}

& TB Acc. (\%) & Pneumonia Acc. (\%) & Normal Acc. (\%) & Overall Acc. (\%) \\
\hline
DenseNet & 95.0 & 97.5 & 97.2 & 96.55 \\
EfficientNet & 93.5 & 96.2 & 96.0 & 95.57 \\
ResNet & 93.2 & 96.0 & 96.3 & 95.51 \\

\end{tabular}
\caption{Acc. Table for models.}
\label{tab:accuracy}
\end{table}

\begin{figure}
    \centering
        \includegraphics[width=0.75\linewidth]{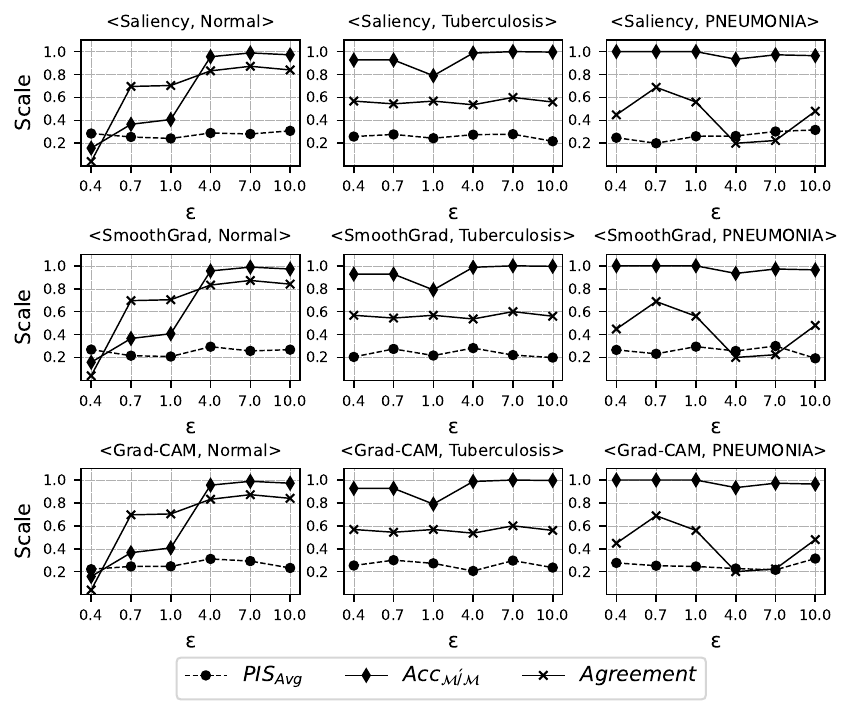}
        \caption{Performance of explainers (ResNet-34)}
        \label{dp-resnet}
        
        \includegraphics[width=0.75\linewidth]{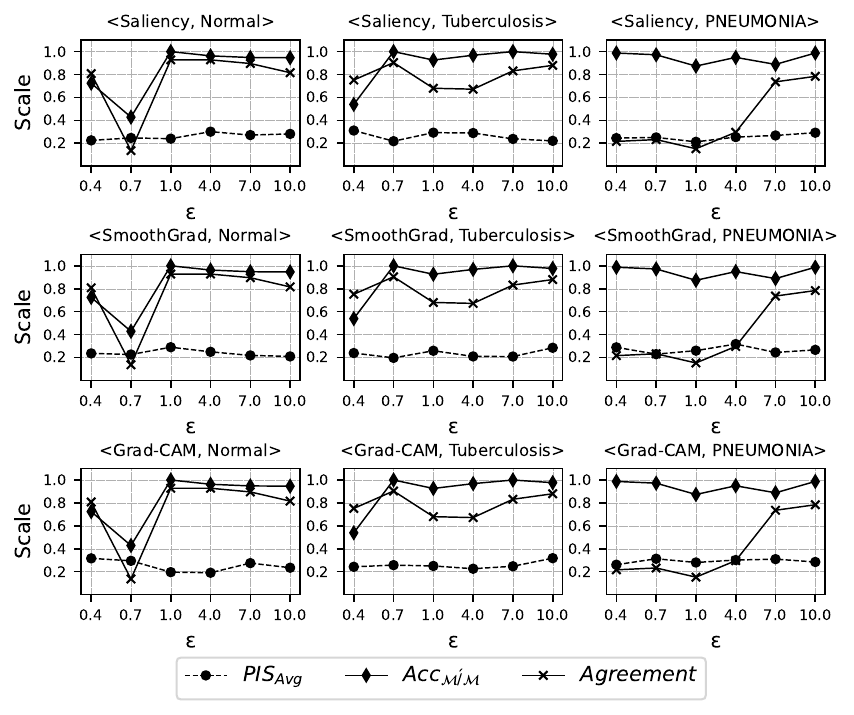}
        \caption{Performance of explainers (DenseNet-121)}
        \label{dp-densenet}

\end{figure}
\begin{figure}
    \centering
        \includegraphics[width=0.75\linewidth]{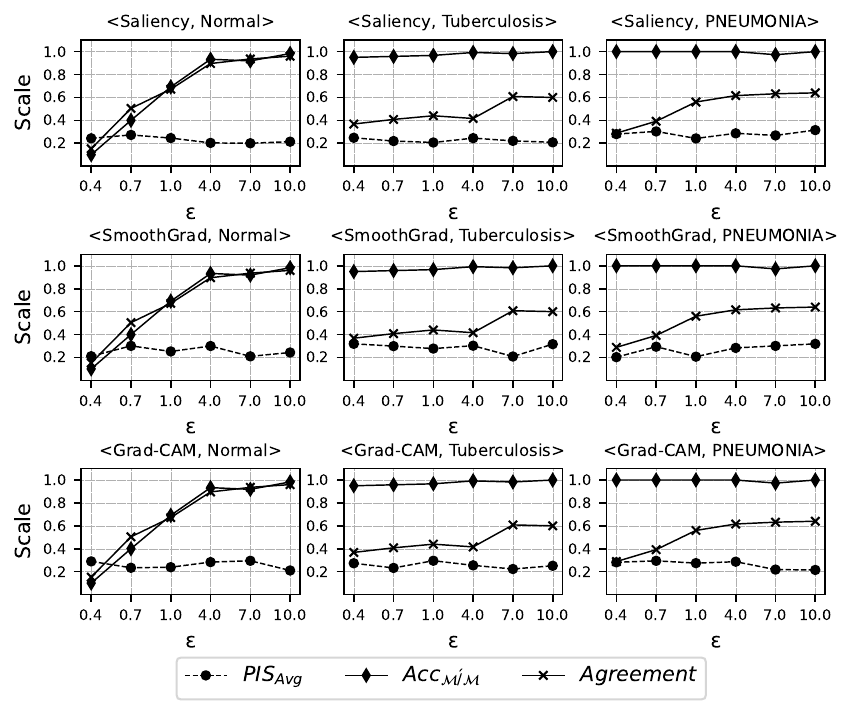}
        \caption{Performance of explainers (EfficientNet-V2)}
        \label{dp-effnet}
    
    % \caption{Performance of the post-hoc explainers for DenseNet-121, ResNet-34, and EfficientNet-v2}
    % \label{dp-all}
\end{figure}

\section{Key Observation}\label{DIS}

In this section, we present our experimental results to find out whether the post-hoc explainers agree with the \texttt{Localization Assumption} (LA), defined in Section~\ref{Sec-LA}. For \texttt{Integrated Gradients} and \texttt{Grad-Shap}, across all models and test samples chosen, we obtained DS of more than 45\%, which directly violates our DS threshold of $15\%$. As we consider DS as a sanity check for LA, and these two explainers fail this test, we exclude them from further discussions. Across models, for around $30-40\%$ of all test samples, \texttt{Grad-Cam} explanations violated the DS threshold of $15\%$; consequently, we have considered the rest of the test samples to calculate PIS for \texttt{Grad-Cam}. All test samples passed the DS test across models for \texttt{Saliency} and \texttt{SmoothGrad}.

We report our findings for the three networks: DenseNet-121, ResNet-34, and EfficientNet-v2 in Figure \ref{dp-densenet}, \ref{dp-resnet}, and \ref{dp-effnet} respectively. We report class-wise results for each of these networks against each of the three explainers that pass the DS test. Against each $\epsilon$, we report average of the PIS values over the test set ($PIS_{Avg}$), $Acc_{\mathcal{M'}/\mathcal{M}}$, and agreement on the ``hard predictions'' obtained from $\mathcal{M}$ and $\mathcal{M'}$ over the test set ($Agreement$). $PIS_{Avg}$ gives the extent to which an explainer agrees with the LA in a particular setup; $Acc_{\mathcal{M'}/\mathcal{M}}$ quantifies the \textit{Perf Comp.} requirement, and $Agreement$ quantifies the \textit{Alignment} requirement for the private model to be \textit{functionally comparable} w.r.t. the non-private counterpart, defined in Section~\ref{Sec-LA}. We observe all of the explainers across $\epsilon$, classes, and models are performing very poorly in terms of $PIS_{Avg}$\footnote{One may argue that what if we measure similarity without segregating attributes? This, although disregards our philosophy on similarly measure, still we did, and to be precise, never found $PIS_{Avg} > 0.16$.}. In fact, the $PIS_{Avg}$ never crosses $0.32$; such a low $PIS_{Avg}$ indicates that none of the explainers agree with the LA. In almost all cases, except for the Normal class against low $\epsilon$ values, $Acc_{\mathcal{M'}/\mathcal{M}}\approx 1$, which indicates that almost all private models achieved near equal accuracy as of their base counterparts. Nearly $70\%$ of cases achieved $Agreement > 0.50$. There is no apparent relation or trend the $PIS_{Avg}$ follows with $Acc_{\mathcal{M'}/\mathcal{M}}$ and $Agreement$. All in all, the explanation quality is consistently poor, no matter what happens to the other parameters. We observe the very same trend for another benchmark dataset we considered (see Appendix \ref{App A}). Empirically, it is clear that all gradient-based explainers disregard LA and thus seem not to be suitable with DP models. With that note, we (partially) answered our primary research question that DP models and off-the-shelf explainers (apparently) \textbf{don't go} \textit{well together}. But why is it so? We'll try to find out in the next section. Also, we'll explore an alternative route to generate private explanations in the subsequent section. 

\section{Why Explainers Performed so \textit{Poorly}?}\label{EXP}
As noted in the previous section, the explanations from private and non-private models are largely uncorrelated. Private models are trained using DP-SGD, which involves gradient clipping and adding calibrated noise during backpropagation to achieve DP. To understand the divergent behavior of explainers between these models, we first analyze how DP training alters model parameters. We conduct a comparative study on the (layerwise) representations learned by both private and non-private models, followed by a rigorous investigation into the (local) sensitivity of these representations. While the former reveals how DP training modifies the parameter space, the latter explains how these modifications influence the model’s reaction to inputs, ultimately affecting explainer behavior. To the best of our knowledge, this is the first study to comprehensively examine the representations learned by DP models and compare them to non-private models. Moreover, the layer-wise sensitivity analysis has not been previously explored for any model, including DP models. We include the implementation details in Appendix C.

 \subsection{Did the models \textit{perceive} the same \textit{information}?} \label{repre}

        In its forward pass, a neural network hierarchically transforms the given input into increasingly abstract (and complex) representations. A Representation (also referred to as a feature representation or simply \{intermediate\} feature) is the set of activations stored layerwise in the network during the forward pass for a given input. We begin our investigation by examining how these representations differ in the private and non-private models, as this reveals how these two different models process and transform the input data across layers. Understanding the alignment of these representations is critical for analyzing how DP training has altered the model's parameters and how these changes influence the model's overall reasoning.

        However, since the training processes for DP model is fundamentally different and designed to achieve distinct goals, we first conduct a formal assessment to determine whether any statistical dependence exists between their layerwise representations. This step ensures that the representations are comparable before further analysis. We denote the representation at layer $l$ as $\sigma_{l}$.

        Statistical tests for activations are often non-trivial and subject to a few constrain,ts such as invariance to permutation of the neurons, orthogonal transformation \citep{klabunde2024similarityneuralnetworkmodels} etc. Consequently, we use the Hilbert-Schmidt Independence Criterion (HSIC) \citep{gretton2007kernel} which operates over a Reproducing Kernel Hilbert Space (RKHS) and can detect dependencies across complex, high-dimensional variables while respecting the aforesaid constraints. We employ the HSIC unconditional independence test using the two-parameter $\gamma$ approximation scheme with a p-value cut-off of $0.05$ \citep{JMLR:v13:gretton12a}. For HSIC, we used both linear and non-linear (RBF) kernels, but obtained similar results.

        From our findings, we were \textbf{able to reject} the null hypothesis $\mathcal{H}_0$ for almost all layers\footnote{For example, across models we never crossed $7-10\%$ of the layers for any $\epsilon$ where we accept $\mathcal{H}_0$}. In other words, across models, layerwise representations are \textbf{not} independent, for all $\epsilon$. Since the representations are sufficiently comparable, we'll now compute their similarity layerwise. However, before a full-fledged comparison, we present a short note on representational similarity (RS).

        \subsubsection{On Representational Similarity}
        Representational similarity measures compare neural networks by computing similarity between activations of a fixed set of inputs (in our case, the test set) at a given pair of layers. There's been an extensive line of work on this domain however, each of them can be categorized depending on the notion of \textit{similarity} it regards. Each similarity measure follows a set of assumptions. In our paper, we have considered the most regarded set of assumptions and its corresponding similarity measure, namely Centered Kernel Alignment (CKA) \citep{kornblith2019similarity, gretton2007kernel}. CKA is built upon HSIC and bounded within the (closed) interval of $[0, 1]$. However, there are a few (discovered) shortcomings of CKA, as discussed below.

        The wildly used CKA with the linear kernel, is equivalent to the RV coefficient and it is already shown that the same seldom yields values close to $1$ due to one of its inherent shortcomings: although the coefficient is inherently constrained to values between 0 and 1, it \textit{rarely} reaches values near 1 because the denominator is typically much larger relative to its theoretical maximum value \citep{puccetti2022measuring}. Also, the less adapted RBF-CKA is highly sensitive to the kernel width and ineffective for small values \citep{davari2022reliability}.

        Furthermore, very recently, Cui et al. \citep{cui2022deconfounded} have discovered that the inter-example (dis)similarity in the representation space works as a confounder. In their words, ``\textit{This leads to spuriously high CKAs even between two random neural networks, and counter-intuitive conclusions when comparing CKAs on sets of models trained on different domains $...$}". They fixed this problem by regressing out the confounder from the similarity matrices of two representations. Our analysis uses this \textit{deconfounded} version of CKA (dCKA).

\begin{figure}[ht]
    \centering
    
        \includegraphics[width=\linewidth]{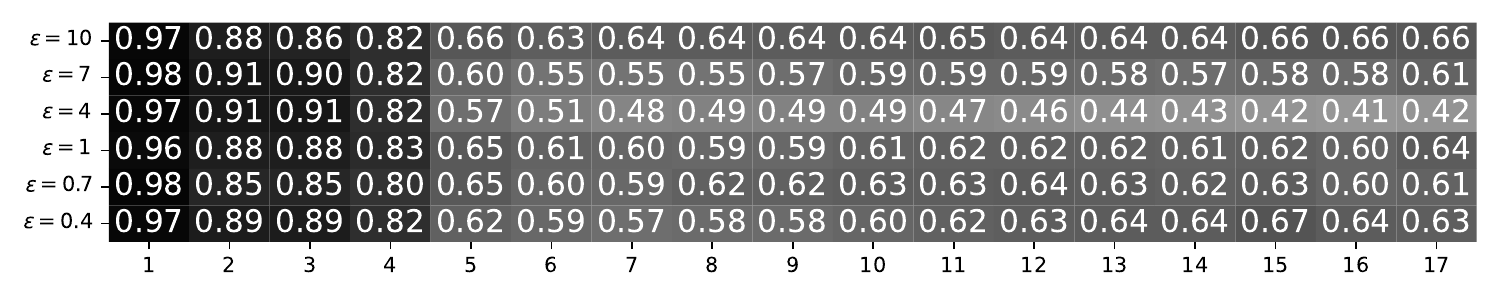}
        \caption{dCKA heatmaps for ResNet-34}
        \label{fig:Investigating_Resnet}
    
    % \hfill
    
        \includegraphics[width=\linewidth]{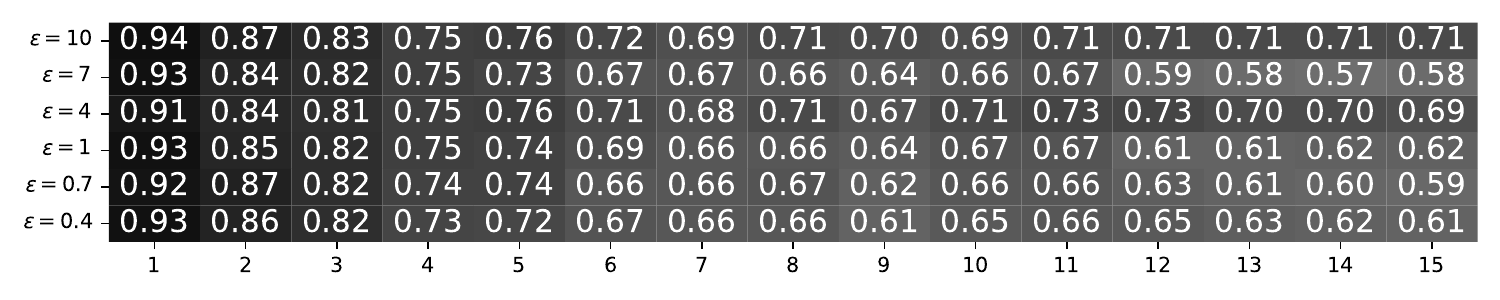}
        \caption{dCKA heatmaps for DenseNet-121}
        \label{fig:Investigating_Densenet}
    
    % \hfill
    
        \includegraphics[width=\linewidth]{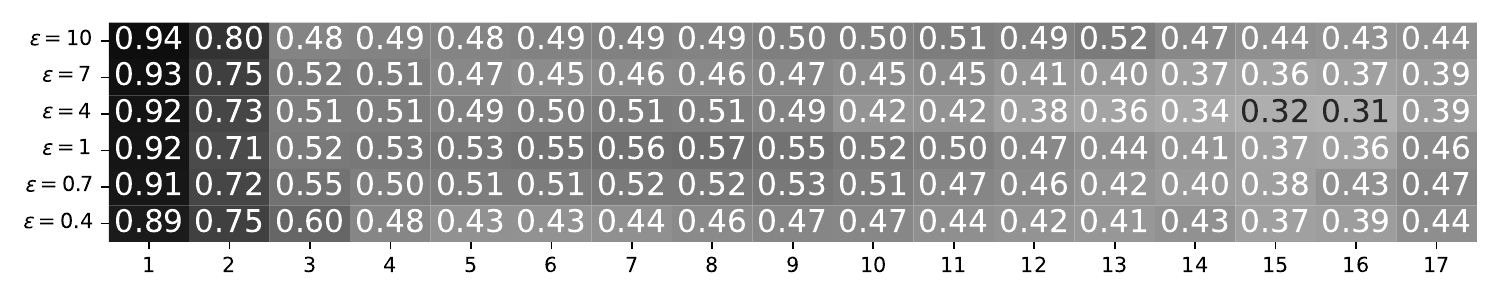}

        \caption{dCKA heatmaps for EfficientNet-V2}
        \label{fig:Investigating_Efficientnet}
    
    \label{fig:Investigating_All}
\end{figure}

        In our approach, since both the non-private and private models share the same architecture, we perform a layer-wise comparison between the corresponding layers of the non-private model and its private counterpart(s). However, as the models under consideration are excessively large, reporting Representational Similarity (RS) for each activation layer individually can become overwhelming due to the sheer volume of data. Instead, depending on the number of layers, we grouped the activation layers into 15/17 clusters\footnote{For ResNet-34 and EfficientNet-V2, we considered all 17 \texttt{ReLU} and 102 \texttt{SiLU} layers respectively, constituting 17 clusters ($17 | 102$). For DenseNet-121, we considered all 120 \texttt{ReLU} layers, constituting 15 clusters ($15 | 120$).} and reported the median RS for each cluster, as shown in Figures \ref{fig:Investigating_Resnet}, \ref{fig:Investigating_Densenet}, and \ref{fig:Investigating_Efficientnet}. From our findings, it is evident that any pair of models learn somewhat similar representations. In the initial layers, they learn almost identical representations, but as we progress through the layers, the similarity deteriorates. Interestingly, the last few layers in all model pairs exhibit significant dissimilarity compared to the first few.

        All in all, the private models don't perceive the data in a strongly correlated manner w.r.t their non-private counterparts. However, we argue that this is not the only reason for the explanations to be altered, as explanations are based on the \textit{sensitivity} of the features subject to a class of interest, the model shows. In the second phase of our investigation, we shall look into how much, for an obtained output, the model is \textit{sensitive} to all the layerwise representations it obtained in the forward pass.

    \subsection{Do the models show \textit{similar} sensitivity?}

In this section, we investigate how sensitive the model’s output is to the layerwise representations generated during the forward pass. Formally, for a given layer \( l \) and a class of interest (here, ``hard prediction'') \( \Theta \), we compute the gradient: \( \nabla_{\sigma_l} \Theta \). This gradient is particularly important for two reasons: first, it quantifies the influence of an infinitesimal perturbation in the representations at layer \( l \) on the final output. Second, as previously discussed in section \ref{xai}, gradient-based explainers also leverage \( \nabla_{\sigma_l} \Theta \) to generate explanations\footnote{Typically, \( l \) refers to the input layer, except for \texttt{Grad-CAM}.}. Thus, explanations, in essence, can be unanimously viewed as the output of a class of \textit{mechanisms} applied to \(\nabla_{\sigma_l}\Theta\), subject to specific input(s), layers and/or baselines (wherever required).

This is why the comparability of explanations across models depends on whether \(\nabla_{\sigma_l}\Theta\) obtained layerwise are comparable between private and non-private models. Moreover, as the level of abstraction in representations varies across layers, we examined \(\nabla_{\sigma_l}\Theta\) at all activation layers to understand the full spectrum of sensitivity across the models.

However, unlike representations, working with \(\nabla_{\sigma_l}\Theta\) presents a few more challenges. First, since it directly depends on the model’s final output, we need to ensure that both models predict the same class for the input instance in order to make the gradients comparable. If the models predict different classes, the gradients will reflect sensitivities toward those different outputs, making direct comparisons inappropriate. To address this, we could restrict our analysis to the subset of the test set where both models make identical predictions. However, this approach has a distinct problem: RS is calculated over the entire test set, whereas \(\nabla_{\sigma_l}\Theta\) would be evaluated over this reduced subset. This mismatch in the sample space (test set) immediately invalidates the direct application of \textit{any} quantitative similarity measures. Conversely, if we use the reduced dataset for evaluating RS, we will not be able to capture the \textit{true} similarity between layers. Additionally, unlike RS, we lack well-established assumptions for defining similarity in the case of \(\nabla_{\sigma_l}\Theta\). These inherent challenges make it hard to establish a one-to-one correspondence between the similarity of layerwise representations and the similarity of their corresponding sensitivity(\(\nabla_{\sigma_l}\Theta\)).

Due to these inherent challenges, we only conduct hypothesis testing using HSIC, as previously mentioned, to check whether the sensitivity of representations is statistically (in)dependent, where the corresponding outputs match\footnote{Note, this is the dataset considered for PIS.}. Here also, we employed both linear and non-linear (RBF) kernels but obtained similar results.

From our findings, we were \textbf{unable to reject} the null hypothesis \(\mathcal{H}_0\) for almost all layers\footnote{In no model did we reject \(\mathcal{H}_0\) for more than 3-7\% of the layers across \(\epsilon\). Notably, for DenseNet-121, all layers were found to be independent}. In other words, unlike representations, the \textit{sensitivity} of representations is independent across models. Regarding \texttt{Grad-CAM}, we typically use the last layer to generate the CAM. Our experiment showed that the gradients of these layers exhibit independence, meaning the CAMs produced by different models will not be comparable. Similarly, other explainers relying on \(\nabla_{\sigma_l}\Theta\) cannot produce aligned and substantially comparable explanations, as representations across levels along with the last layer show \textbf{independent} sensitivity between non-private and their private counterparts\footnote{A natural question arises: could an explainer, despite of consistently incomparable sets of \(\nabla_{\sigma_l}\Theta\), produce similar explanations? Based on our experiments, we did not identify any commonly used explainer that exhibited this property for the models selected. Furthermore, we argue on the \texttt{faithfulness} of such explainers, if they exist, would be questionable. Conversely, assessing the quality of explanations when (local) sensitivity is consistently comparable across \textit{any} pair of models falls outside the scope of this paper.
}.

$\bigodot$ Overall, neither the DP models extract the features the way a non-private model does, nor can they exhibit the sensitivity over the features in a similar fashion. $\nabla_{\sigma_{l}}{\Theta}$ can be viewed as how $\Theta$ changes for a \textit{infinitesimally} small perturbation around $\sigma_{l}$. We have observed $\sigma_{l}$ are mildly similar across layers for non-private and private models; however, `$\nabla$' operator being a crude first-order approximation only addresses the immediate, linear response of the output w.r.t perturbations around $\sigma_{l}$ which, in our case, is \textbf{independent} to that of non-private model(s). In other words, the \textit{different} set of parameters obtained with DP training is \textbf{sufficient} for the non-private model to show divergent sensitivity across layers, which, in turn, makes the explanations incomparable. DP training was primarily meant to resist MIA w.r.t. the training data, but the training goes much beyond the scope, and drastically alters the overall representation and sensitivity space of a model. Which, as demonstrated, makes the popular post-hoc methods fundamentally nonfunctional. With that note, we now have arrived at a \textbf{firm conclusion} that due to DP models' inherent nature, off-the-shelf explainers \textbf{don't go} \textit{together} with them.

\section{Is there any alternative way to have private explanations?}\label{LDP-1}
We now know that the DP models cannot accommodate widely used post-hoc explainers primarily due to their own nature. As a result, we cannot move forward with DP models to obtain private explanations that will be useful as a proxy to that of the non-DP counterpart. However, \textit{just} for private explanations, we don't need that. We opt for an \textit{alternative route} to achieve the same \textit{locally}.

In this approach, rather than training the model with DP-SGD, we use the non-private model as usual and take its explanations. We add calibrated noise to the explanation to make it Local Differential Private (LDP)\newline

$\blacksquare$ Formally, given a function $f : D \to \mathbb{R}^d$, the Laplace mechanism $\mathcal{A}$ is defined as:
\[
\mathcal{A}(D) = f(D) + \text{Lap}(0|t)^d
\],

Where $\mathcal{A}$ satisfies $\epsilon$-differential privacy for $t = \frac{\Delta f}{\epsilon}$, and $\Delta f$ is the sensitivity\footnote{This sensitivity is different from the (local) sensitivity we discussed in section \ref{EXP}.} of $f$ \citep{dwork2014algorithmic}.

Here, noise is directly added to the heatmap (explanation) obtained from the explainer. Following Fan \citep{fan2018image}, the sensitivity of a query for an image with \(c\) channels and \(k\) possible pixel intensities is \((k-1)nc\), where \(n\) is the \textit{maximum} number of \textit{differing pixels}. To ensure \(\epsilon\)-differential privacy, if we apply the Laplace mechanism to each pixel in each channel, scaling by \((k - 1)nc / \epsilon\), such noise can obscure image semantics, particularly for smaller \(\epsilon\). To mitigate this, Fan \citep{fan2018image} advises dividing the image into \(b \times b\) grids, averaging each grid's pixels, which reduces the sensitivity ($\Delta f$) to \((k-1)nc / b^2\). Therefore, for images (including heatmaps) with a single channel, the sensitivity becomes \(255n / b^2\). We empirically tested with a diverse set of $(n, b)$ for LDP explanations and communicated the results with the concerned physicians. We get the best response for the tuple (\(16, 14\)), and the most competent explainers were \texttt{Grad-Shap} and \texttt{Integrated Gradients}. In our case, we get useful results $\epsilon = 4$ onwards. We selected the ResNet-34 to demonstrate our results for all four explainers, which can be found in the supplementary material.
 
\subsection{What Could Be the Notion of PIS and LA Here?}
After \textit{LDP-fying} with $b \times b$ grids, pixel-wise comparisons are impractical. However, the postulates of LA does hold here as well: the most competent explanations should be least affected by LDP transformations. Therefore, such private explanations ($s'$) should be most \textit{similar} to their non-private counterpart (\textit{s}) and serve as a close proxy. To quantify the degradation of the \textit{quality} of $s'$ LDP, we employ the Structural Similarity Index (SSIM) \citep{wang2004image} as PIS in this context. In the absence of pixel-by-pixel comparison, unlike global DP, SSIM aims to reflect how much essential information is retained post-\textit{LDP}fication, primarily focusing on several perceptual similarities by evaluating structural components in $s'$. However, as balancing the degree of noise to achieve a desired privacy level may severely affect the structural fidelity of $s'$, thus inevitably diminishes SSIM, we primarily use SSIM evaluation as an \textit{elimination test} to discard any obfuscated \textit{LDP}-fied explanations (having substantially low, near zero, or negative SSIM). Furthermore, we \textit{eliminated} \texttt{Grad-CAM} as its small region of interests ($7 \times 7$ or $8 \times 8$) tend to get overwhelmed by \textit{LDP} noise, and fixing the new set of hyperparameters here is also challenging. We roughly get PIS between $0.4 - 0.5$ for all other explainers (Table \ref{t2}); however, the concerned physician recommended \texttt{Integrated Gradients} and \texttt{Grad-Shap} to be \textit{most useful} throughout our experiments.

\begin{table}[h!]
\centering
\resizebox{\linewidth}{!}{
\begin{tabular}{c|c|c|c|c}
 & \text{Grad-SHAP} & \text{Integrated Gradient} & \text{Saliency} & \text{SmoothGrad} \\
 \hline
\text{Tuberculosis} & 0.52 & 0.50 & 0.49 & 0.52 \\
\hline
\text{Pneumonia} & 0.53 & 0.51 & 0.43 & 0.49 \\
\hline
\text{Normal} & 0.56 & 0.54 & 0.49 & 0.53 \\
\end{tabular}
}
\caption{Mean SSIM for \textit{LDP}-fied explanations.}
\label{t2}
\end{table}

In this setup, the model is non-private, but the explanations are private; we name this setup as Hybrid DP. Our novel software is outlined using this setup in Figure \ref{software}.

$\bigodot$ To summarize:
\begin{table}[h]
\centering
\resizebox{\linewidth}{!}{%
\begin{tabular}{c|c|c|c|c|c}
& Drop in Accuracy & Private Model & Private Output & Private Explanation & Useful Explanation \\
\hline
Global DP & \checkmark & \checkmark & \checkmark & \checkmark & $\times$ \\
\hline
Hybrid DP & $\times$ & $\times$ & $\times$ & \checkmark & \checkmark \\
\end{tabular}}
\caption{Comparison of Privacy Methods}
\end{table}

\section{The Novel Software Pipeline}\label{swe}

So far, we have investigated non-trivial intricacies of DP models and XAI methods, and we \textit{figured out} an alternative way to generate private explanations using LDP. Based on all the insights we have consolidated so far, now we propose a privacy-preserving software pipeline that aims to reconcile model explainability and privacy\footnote{This proposed Software Pipeline is \textit{exclusively} based upon our exhaustive findings and insights that we have elaborated throughout our study, we do not claim this to be the (only) ideal setup for high-stakes.}.

\textbf{System Overview and Security Measures}: At the system’s entry point, all incoming medical images pass through a trained autoencoder (AE) \citep{neloy2024comprehensive}, which filters out anomalous inputs. We achieve $94\%$ mean accuracy on anomaly detection at a reconstruction loss threshold ($\kappa = 0.07$), using randomly sampled 50 test images from each non-target class (Cardiomegaly, Aortic enlargement from \citep{nguyen2022vindr} dataset) as anomalous examples. Anomalous sample classification can be done with various methods, but we chose AE due to its widespread use in anomaly detection \citep{neloy2024comprehensive}.

\begin{figure}[h!]
    \centering
    \includegraphics[width=\linewidth]{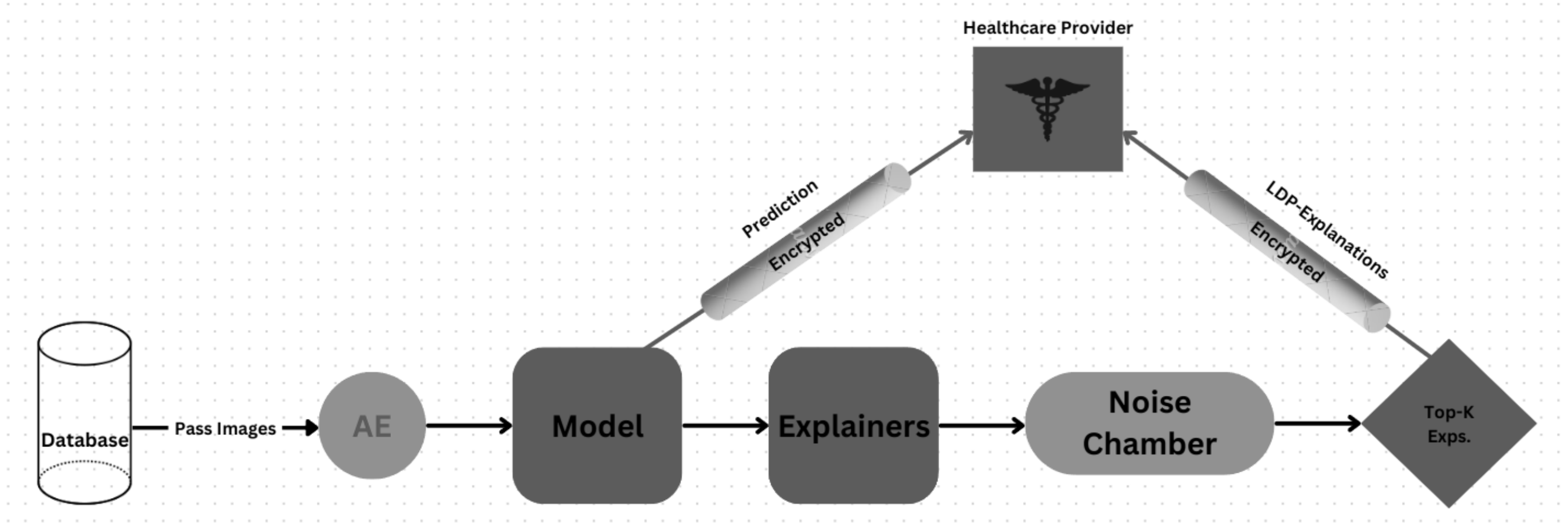}
    \caption{Outline of the Software}
    \label{software}
\end{figure}

\textbf{Core Processing Pipeline}: Validated images go through the core processing pipeline, where a non-private model makes predictions. The model's prediction (only the label) and the LDP-fied explanation (described in Section \ref{LDP-1}) are generated. These explanations are assessed for quality using SSIM (as shown in Figure \ref{software}). Explanations with poor SSIM scores (e.g., negative or near-zero values) are discarded to avoid wasting the physician’s time.

The remaining top-K highest-quality LDP-fied explanations (here, K = 1-2), along with the model's prediction, are securely transmitted to the physician via an encrypted channel. This information is sent to the physician(s), not the patient, at this moment.

\textbf{Final Communication Protocol}: The final prediction and explanation, upon physicians' approval, are communicated to the patient.

$\blacksquare$ \textsc{How much `Compromise' is required to `Reconcile'?}
In our proposed software pipeline, we refute DP models for reconciling privacy and explainability in high-stakes, as XAI methods and DP models don't \textit{go together}. Therefore, we must clearly \textit{compromise} as we have to fall back to \textit{non-private} model when we aim to gracefully \textit{balance} both privacy and explainability in high-stakes. With LDP-fied explanations, we mitigate the privacy breach from explanation (as mentioned in Section \ref{TPAB}) and we do not share the prediction vector, but only the label, to protect against MIA attacks from it. However, we acknowledge that our proposed pipeline is inappropriate for systems with a sheer requirement of both prediction vector and \textit{useful} private explanations. From our findings, it is clear that, at least with DP models, building such systems is not possible. One may argue that if we LDP-fy the prediction vector coming from a non-private model by adding calibrated noise, it could be a \textit{better} option than discarding the non-private prediction vector. We do not advocate making the prediction vector LDP by injecting noise, as excessive noise could alter the \textit{actual} prediction and/or mislead decision-makers with false prediction confidence, impairing both model interpretability and utility. We note that sharing the model's prediction (only the label) is the least possible amount of \textit{privacy risk} we take to make our pipeline \textit{functional}. Our findings show that reconciling privacy and explainability in high-stakes situations comes with several other \textit{silent factors} that need to be exclusively addressed, and is non-trivial. While this study aims to serve as a rigorous starting point, canonically reconciling privacy with explainability, with \textit{minimum compromise}, is still an open challenge that demands further investigation, probably even beyond the setup and constraints we discussed in this paper.

\section{Related Work}
\label{RELW}
\textit{Privacy-preserving machine learning (PPML)} and \textit{Explainable AI (XAI)} are well-studied research areas, and we've outlined the topics under consideration for this study in section \ref{pril}. For a comprehensive review of \textit{PPML} and \textit{XAI}, we direct readers to the excellent surveys by Boulemtafes et al. \citep{boulemtafes2020review} and Saeed et al. \citep{saeed2023explainable}, respectively. 

\textit{Privacy-preserving explainable AI (PPXAI).} PPXAI methods are primarily an emerging class of XAI methods specifically made to protect \textit{sensitive information} in their explanation (desirably with provable guarantees), primarily about the training data. PPXAI methods have only recently begun to emerge, with approaches such as differentially private Locally Linear Maps (LLM) \citep{harder2020interpretable}, differentially private feature-based model explanations \citep{patel2022model}, and differentially private counterfactual explanations \citep{mochaourab2021robust, yang2022differentially}, etc. Harder et al. in \citep{harder2020interpretable} proposed a family of simple models to \textit{approximate} complex models using locally linear maps for each class, which aims to provide DP explanations. Patel et al. \citep{patel2022model} introduced a perturbation-based algorithm that aims to make the \textit{explanation dataset} DP, using which they, in a \textit{black-box} fashion, generate a private explanation for a given \textit{point-of-interest}. Mochaourab
et al. \citep{mochaourab2021robust} made a DP Support Vector Machine
and introduce methods for generating \textit{robust} counterfactuals, Yang et al. \citep{yang2022differentially} utilises a DP autoencoder that creates privacy-preserving prototypes by perturbing input data to minimize distance to a counterfactual, while promoting a specific class outcome. For a detailed overview of PPXAI, we direct readers to the comprehensive survey by Nguyen et al. \citep{nguyen2024surveyprivacypreservingmodelexplanations}. 

Our work is completely orthogonal to PPXAI methods. It is worth noting that our work doesn't propose or analyse \textit{any} PPXAI method but explains the interplay between off-the-shelf, well-regarded explainers with DP models. We first establish the desiderata for DP explanations and then mechanistically explore why DP models are particularly challenging to explain using traditional post-hoc explainers. We not only investigate global DP and its inherent limitations for RTE but also explore an alternative route to generate \textit{useful} private explanations using local DP to generate proxy private explanations.

Recently, researchers have begun investigating the quality of explanations in privacy-preserving environments. Bozorgpanah et al.~\citep{bozorgpanah2022privacy} generated private datasets (benchmark tabular datasets) using masking methods and compared the Shapley values \citep{lundberg2017unified} for test instances given the two models: trained on non-private data and trained on private data. They found minimal effect on moderate protection. Lucieri et al.~\citep{lucieri2023translating} investigated the effect of DP training on concept-based explanations on biomedical datasets, and found that DP decreases average Concept Activation Vectors (CAV) accuracy and increases standard deviation; whereas \citep{saifullah2024privacy} has reported that differential privacy and federated learning may yield `noisy' feature attribution scores for post-hoc explainers. Berning et al.~\citep{berning2024trade} found that k-anonymity degrades the quality of counterfactual explanations on a tabular dataset. Bozorgpanah et al.~\citep{bozorgpanah2024explainable} generated private datasets (benchmark tabular datasets) using masking methods and noise addition and applied TreeSHAP to achieve plausible explanations. 

Our work shares some similarities with the aforementioned studies, but it differs significantly in approach. While previous research typically uses quality measures from the XAI literature in an ad-hoc manner within privacy-preserving environments, our systematic investigation is grounded in our own proposed postulate and measures: the Localization Assumption (LA) and Privacy Invariance Score (PIS). While earlier works focus on measuring the \textit{degradation} in explanation quality within privacy-preserving settings using a limited set of explainers, we demonstrate under a tightly controlled environment that the issue isn't simply any \textit{degradation} in explanation quality. Rather, it's the intrinsic nature of DP models themselves that leads to disobeying LA and producing orthogonal explanations across all commonly used gradient-based explainers, and we are the first ones to mechanistically interpret this very phenomenon. Moreover, many prior studies select explainers and quality measures in a rather ad-hoc fashion. In contrast, we acknowledge our apprehension regarding the suitability of model-agnostic explainers for DP models. Additionally, while Lucieri et al. \citep{lucieri2023translating} and Saifullah et al. \citep{saifullah2024privacy} have focused on global DP, the majority of existing research has dealt with local DP.

\textit{Analyzing network similarity.} Analyzing neural network similarity is crucial for interpreting and improving model behavior. This can be broadly classified into two main types: (i) representational similarity (RS), which measures differences in intermediate layer activations, and (ii) functional similarity, which evaluates discrepancies in model outputs \citep{klabunde2024similarityneuralnetworkmodels}.

To investigate why DP models fail to accommodate common post-hoc explainers, we investigated RS to examine layer-wise changes in activations and their sensitivities. However, in such cases, the similarity measures researchers employ adhere to specific invariances: Permutations, Orthogonal Transformations, Isotropic Scaling, Translations, etc, are to name a few. We have considered the most commonly used set of invariances and chose CKA for measuring RS, but CKA, by its design, comes with a few shortcomings as discussed in section \ref{EXP}. Consequently, we chose dCKA proposed by Cui et al. \citep{cui2022deconfounded} for our analysis. For a broader view on neural network similarity, we refer the readers to the wonderful survey by Klabunde et al. \citep{klabunde2024similarityneuralnetworkmodels}.

\section{Conclusion and Future Studies}
\label{CONC}
We started our research with a simple question of whether we can achieve RTP and RTE \textit{together}. We kept an eye on the pitfalls of commonly used evaluation metrics for explainability and proposed our desiderata, and found that no commonly used gradient-based explainers are useful for private models. We investigate the activations inside a DP model and how the model is sensitive towards those; we discovered that the intrinsic behaviour of DP models is the key reason behind this. Our \textit{mechanistic} insights of the private models across different privacy guarantees highlight how DP training alters the internal representations and their sensitivities. It gives a fresh perspective on interpreting DP models and their behaviour from a nuanced angle. Lastly, we make use of LDP to achieve private explanations and conclude our study by outlining the pipeline for the industrial software for our use case that respects both RTP and RTE.

In future studies, we aim to go deeper into the sensitivity landscape of DP models by investigating second-order and higher-order derivatives of the model's output w.r.t. representations, which may be helpful while investigating a few niche explainers' behaviour (e.g.: Integrated Hessian \citep{janizek2021explaining}), which doesn't directly leverage \textit{immediate} first-order gradients. However, such explainers are still not widely used in high-stakes unlike the ones we have considered in this paper. While this work elucidates the interplay between the highly regarded gradient-based explainers and DP models, and \textit{explains} the degradation of explanations of DP models,   examining higher-order derivatives could potentially uncover richer structural patterns in the sensitivity space that aren't immediately apparent through first-order analysis.

\section*{Acknowledgement}
We are thankful to Jim Conant for the fruitful discussion on high-dimensional geometry. We are thankful to Yogesh Kumar, the co-author of dCKA~\citep{cui2022deconfounded}, for a productive discussion. We are also thankful to Dr. Venkata Abhinay Talasila from SRM University AP for his immense cooperation throughout.

\bibliographystyle{plainnat}
\bibliography{draft}

\begin{thebibliography}{109}
\providecommand{\natexlab}[1]{#1}
\providecommand{\url}[1]{\texttt{#1}}
\expandafter\ifx\csname urlstyle\endcsname\relax
  \providecommand{\doi}[1]{doi: #1}\else
  \providecommand{\doi}{doi: \begingroup \urlstyle{rm}\Url}\fi

\bibitem[Abadi et~al.(2016)Abadi, Chu, Goodfellow, McMahan, Mironov, Talwar, and Zhang]{Abadi_2016}
Martin Abadi, Andy Chu, Ian Goodfellow, H.~Brendan McMahan, Ilya Mironov, Kunal Talwar, and Li~Zhang.
\newblock Deep learning with differential privacy.
\newblock In \emph{Proceedings of the 2016 ACM SIGSAC Conference on Computer and Communications Security}, CCS’16. ACM, October 2016.
\newblock \doi{10.1145/2976749.2978318}.
\newblock URL \url{http://dx.doi.org/10.1145/2976749.2978318}.

\bibitem[Adebayo et~al.(2018)Adebayo, Gilmer, Muelly, Goodfellow, Hardt, and Kim]{adebayo2018sanity}
Julius Adebayo, Justin Gilmer, Michael Muelly, Ian Goodfellow, Moritz Hardt, and Been Kim.
\newblock Sanity checks for saliency maps.
\newblock \emph{Advances in neural information processing systems}, 31, 2018.

\bibitem[Al-qaness et~al.(2024)Al-qaness, Zhu, AL-Alimi, Dahou, Alsamhi, Abd~Elaziz, and Ewees]{al2024chest}
Mohammed~AA Al-qaness, Jie Zhu, Dalal AL-Alimi, Abdelghani Dahou, Saeed~Hamood Alsamhi, Mohamed Abd~Elaziz, and Ahmed~A Ewees.
\newblock Chest x-ray images for lung disease detection using deep learning techniques: A comprehensive survey.
\newblock \emph{Archives of Computational Methods in Engineering}, pages 1--35, 2024.

\bibitem[Alvarez-Melis and Jaakkola(2018)]{alvarezmelis2018robustnessinterpretabilitymethods}
David Alvarez-Melis and Tommi~S. Jaakkola.
\newblock On the robustness of interpretability methods, 2018.
\newblock URL \url{https://arxiv.org/abs/1806.08049}.

\bibitem[Ancona et~al.(2017)Ancona, Ceolini, {\"O}ztireli, and Gross]{ancona2017towards}
Marco Ancona, Enea Ceolini, Cengiz {\"O}ztireli, and Markus Gross.
\newblock Towards better understanding of gradient-based attribution methods for deep neural networks.
\newblock \emph{arXiv preprint arXiv:1711.06104}, 2017.

\bibitem[Bach et~al.(2015)Bach, Binder, Montavon, Klauschen, M{\"u}ller, and Samek]{bach2015pixel}
Sebastian Bach, Alexander Binder, Gr{\'e}goire Montavon, Frederick Klauschen, Klaus-Robert M{\"u}ller, and Wojciech Samek.
\newblock On pixel-wise explanations for non-linear classifier decisions by layer-wise relevance propagation.
\newblock \emph{PloS one}, 10\penalty0 (7):\penalty0 e0130140, 2015.

\bibitem[Beltran et~al.(2024)Beltran, Tobaben, J{\"a}lk{\"o}, Loppi, and Honkela]{beltran2024towards}
Sebastian~Rodriguez Beltran, Marlon Tobaben, Joonas J{\"a}lk{\"o}, Niki Loppi, and Antti Honkela.
\newblock Towards efficient and scalable training of differentially private deep learning.
\newblock \emph{arXiv preprint arXiv:2406.17298}, 2024.

\bibitem[Berning et~al.(2024)Berning, Dunning, Spagnuelo, Veugen, and van~der Waa]{berning2024trade}
Sjoerd Berning, Vincent Dunning, Dayana Spagnuelo, Thijs Veugen, and Jasper van~der Waa.
\newblock The trade-off between privacy \& quality for counterfactual explanations.
\newblock In \emph{Proceedings of the 19th International Conference on Availability, Reliability and Security}, pages 1--9, 2024.

\bibitem[Bhatt et~al.(2020)Bhatt, Xiang, Sharma, Weller, Taly, Jia, Ghosh, Puri, Moura, and Eckersley]{bhatt2020explainablemachinelearningdeployment}
Umang Bhatt, Alice Xiang, Shubham Sharma, Adrian Weller, Ankur Taly, Yunhan Jia, Joydeep Ghosh, Ruchir Puri, José M.~F. Moura, and Peter Eckersley.
\newblock Explainable machine learning in deployment, 2020.
\newblock URL \url{https://arxiv.org/abs/1909.06342}.

\bibitem[Blanco-Justicia et~al.(2022)Blanco-Justicia, Sánchez, Domingo-Ferrer, and Muralidhar]{Blanco_Justicia_2022}
Alberto Blanco-Justicia, David Sánchez, Josep Domingo-Ferrer, and Krishnamurty Muralidhar.
\newblock A critical review on the use (and misuse) of differential privacy in machine learning.
\newblock \emph{ACM Computing Surveys}, 55\penalty0 (8):\penalty0 1–16, December 2022.
\newblock ISSN 1557-7341.
\newblock \doi{10.1145/3547139}.
\newblock URL \url{http://dx.doi.org/10.1145/3547139}.

\bibitem[Bora et~al.(2024)Bora, Raja, Terh{\"o}rst, Veldhuis, and Ramachandra]{bora2024why}
Revoti~Prasad Bora, Kiran Raja, Philipp Terh{\"o}rst, Raymond Veldhuis, and Raghavendra Ramachandra.
\newblock Why sanity check for saliency metrics fails?, 2024.
\newblock URL \url{https://openreview.net/forum?id=Pev2ufTzMv}.

\bibitem[Boulemtafes et~al.(2020)Boulemtafes, Derhab, and Challal]{boulemtafes2020review}
Amine Boulemtafes, Abdelouahid Derhab, and Yacine Challal.
\newblock A review of privacy-preserving techniques for deep learning.
\newblock \emph{Neurocomputing}, 384:\penalty0 21--45, 2020.

\bibitem[Bozorgpanah and Torra(2024)]{bozorgpanah2024explainable}
Aso Bozorgpanah and Vicen{\c{c}} Torra.
\newblock Explainable machine learning models with privacy.
\newblock \emph{Progress in Artificial Intelligence}, 13\penalty0 (1):\penalty0 31--50, 2024.

\bibitem[Bozorgpanah et~al.(2022)Bozorgpanah, Torra, and Aliahmadipour]{bozorgpanah2022privacy}
Aso Bozorgpanah, Vicen{\c{c}} Torra, and Laya Aliahmadipour.
\newblock Privacy and explainability: The effects of data protection on shapley values.
\newblock \emph{Technologies}, 10\penalty0 (6):\penalty0 125, 2022.

\bibitem[Buhrmester et~al.(2021)Buhrmester, M{\"u}nch, and Arens]{buhrmester2021analysis}
Vanessa Buhrmester, David M{\"u}nch, and Michael Arens.
\newblock Analysis of explainers of black box deep neural networks for computer vision: A survey.
\newblock \emph{Machine Learning and Knowledge Extraction}, 3\penalty0 (4):\penalty0 966--989, 2021.

\bibitem[Chalasani et~al.(2020)Chalasani, Chen, Chowdhury, Wu, and Jha]{chalasani2020concise}
Prasad Chalasani, Jiefeng Chen, Amrita~Roy Chowdhury, Xi~Wu, and Somesh Jha.
\newblock Concise explanations of neural networks using adversarial training.
\newblock In \emph{International Conference on Machine Learning}, pages 1383--1391. PMLR, 2020.

\bibitem[Chang et~al.(2018)Chang, Creager, Goldenberg, and Duvenaud]{chang2018explaining}
Chun-Hao Chang, Elliot Creager, Anna Goldenberg, and David Duvenaud.
\newblock Explaining image classifiers by counterfactual generation.
\newblock \emph{arXiv preprint arXiv:1807.08024}, 2018.

\bibitem[Cui et~al.(2022)Cui, Kumar, Marttinen, and Kaski]{cui2022deconfounded}
Tianyu Cui, Yogesh Kumar, Pekka Marttinen, and Samuel Kaski.
\newblock Deconfounded representation similarity for comparison of neural networks.
\newblock \emph{Advances in Neural Information Processing Systems}, 35:\penalty0 19138--19151, 2022.

\bibitem[Dasgupta et~al.(2022)Dasgupta, Frost, and Moshkovitz]{dasgupta2022framework}
Sanjoy Dasgupta, Nave Frost, and Michal Moshkovitz.
\newblock Framework for evaluating faithfulness of local explanations.
\newblock In \emph{International Conference on Machine Learning}, pages 4794--4815. PMLR, 2022.

\bibitem[Davari et~al.(2022)Davari, Horoi, Natik, Lajoie, Wolf, and Belilovsky]{davari2022reliability}
MohammadReza Davari, Stefan Horoi, Amine Natik, Guillaume Lajoie, Guy Wolf, and Eugene Belilovsky.
\newblock Reliability of cka as a similarity measure in deep learning.
\newblock \emph{arXiv preprint arXiv:2210.16156}, 2022.

\bibitem[Dong et~al.(2021)Dong, Wang, and Abbas]{dong2021survey}
Shi Dong, Ping Wang, and Khushnood Abbas.
\newblock A survey on deep learning and its applications.
\newblock \emph{Computer Science Review}, 40:\penalty0 100379, 2021.

\bibitem[Dwork et~al.(2014)Dwork, Roth, et~al.]{dwork2014algorithmic}
Cynthia Dwork, Aaron Roth, et~al.
\newblock The algorithmic foundations of differential privacy.
\newblock \emph{Foundations and Trends{\textregistered} in Theoretical Computer Science}, 9\penalty0 (3--4):\penalty0 211--407, 2014.

\bibitem[E.~Ihongbe et~al.(2024)E.~Ihongbe, Fouad, F.~Mahmoud, Rajasekaran, and Bhatia]{e2024evaluating}
Izegbua E.~Ihongbe, Shereen Fouad, Taha F.~Mahmoud, Arvind Rajasekaran, and Bahadar Bhatia.
\newblock Evaluating explainable artificial intelligence (xai) techniques in chest radiology imaging through a human-centered lens.
\newblock \emph{Plos one}, 19\penalty0 (10):\penalty0 e0308758, 2024.

\bibitem[Fan(2018)]{fan2018image}
Liyue Fan.
\newblock Image pixelization with differential privacy.
\newblock In \emph{Data and Applications Security and Privacy XXXII: 32nd Annual IFIP WG 11.3 Conference, DBSec 2018, Bergamo, Italy, July 16--18, 2018, Proceedings 32}, pages 148--162. Springer, 2018.

\bibitem[Fefferman et~al.(2016)Fefferman, Mitter, and Narayanan]{fefferman2016testing}
Charles Fefferman, Sanjoy Mitter, and Hariharan Narayanan.
\newblock Testing the manifold hypothesis.
\newblock \emph{Journal of the American Mathematical Society}, 29\penalty0 (4):\penalty0 983--1049, 2016.

\bibitem[Fioretto et~al.(2022)Fioretto, Tran, Van~Hentenryck, and Zhu]{fioretto2022differential}
Ferdinando Fioretto, Cuong Tran, Pascal Van~Hentenryck, and Keyu Zhu.
\newblock Differential privacy and fairness in decisions and learning tasks: A survey.
\newblock \emph{arXiv preprint arXiv:2202.08187}, 2022.

\bibitem[Fredrikson et~al.(2015)Fredrikson, Jha, and Ristenpart]{fredrikson2015model}
Matt Fredrikson, Somesh Jha, and Thomas Ristenpart.
\newblock Model inversion attacks that exploit confidence information and basic countermeasures.
\newblock In \emph{Proceedings of the 22nd ACM SIGSAC conference on computer and communications security}, pages 1322--1333, 2015.

\bibitem[Gretton et~al.(2007)Gretton, Fukumizu, Teo, Song, Sch{\"o}lkopf, and Smola]{gretton2007kernel}
Arthur Gretton, Kenji Fukumizu, Choon Teo, Le~Song, Bernhard Sch{\"o}lkopf, and Alex Smola.
\newblock A kernel statistical test of independence.
\newblock \emph{Advances in neural information processing systems}, 20, 2007.

\bibitem[Gretton et~al.(2012)Gretton, Borgwardt, Rasch, Sch{{\"o}}lkopf, and Smola]{JMLR:v13:gretton12a}
Arthur Gretton, Karsten~M. Borgwardt, Malte~J. Rasch, Bernhard Sch{{\"o}}lkopf, and Alexander Smola.
\newblock A kernel two-sample test.
\newblock \emph{Journal of Machine Learning Research}, 13\penalty0 (25):\penalty0 723--773, 2012.
\newblock URL \url{http://jmlr.org/papers/v13/gretton12a.html}.

\bibitem[Harder et~al.(2020)Harder, Bauer, and Park]{harder2020interpretable}
Frederik Harder, Matthias Bauer, and Mijung Park.
\newblock Interpretable and differentially private predictions.
\newblock In \emph{Proceedings of the AAAI Conference on Artificial Intelligence}, volume~34, pages 4083--4090, 2020.

\bibitem[Hase et~al.(2021)Hase, Xie, and Bansal]{hase2021out}
Peter Hase, Harry Xie, and Mohit Bansal.
\newblock The out-of-distribution problem in explainability and search methods for feature importance explanations.
\newblock \emph{Advances in neural information processing systems}, 34:\penalty0 3650--3666, 2021.

\bibitem[Haug et~al.(2021)Haug, Z{\"u}rn, El-Jiz, and Kasneci]{haug2021baselines}
Johannes Haug, Stefan Z{\"u}rn, Peter El-Jiz, and Gjergji Kasneci.
\newblock On baselines for local feature attributions.
\newblock \emph{arXiv preprint arXiv:2101.00905}, 2021.

\bibitem[He et~al.(2016)He, Zhang, Ren, and Sun]{he2016deep}
Kaiming He, Xiangyu Zhang, Shaoqing Ren, and Jian Sun.
\newblock Deep residual learning for image recognition.
\newblock In \emph{Proceedings of the IEEE conference on computer vision and pattern recognition}, pages 770--778, 2016.

\bibitem[Hedstr{\"o}m et~al.(2023)Hedstr{\"o}m, Weber, Krakowczyk, Bareeva, Motzkus, Samek, Lapuschkin, and H{\"o}hne]{hedstrom2023quantus}
Anna Hedstr{\"o}m, Leander Weber, Daniel Krakowczyk, Dilyara Bareeva, Franz Motzkus, Wojciech Samek, Sebastian Lapuschkin, and Marina M-C H{\"o}hne.
\newblock Quantus: An explainable ai toolkit for responsible evaluation of neural network explanations and beyond.
\newblock \emph{Journal of Machine Learning Research}, 24\penalty0 (34):\penalty0 1--11, 2023.

\bibitem[Hu et~al.(2022)Hu, Salcic, Sun, Dobbie, Yu, and Zhang]{hu2022membership}
Hongsheng Hu, Zoran Salcic, Lichao Sun, Gillian Dobbie, Philip~S Yu, and Xuyun Zhang.
\newblock Membership inference attacks on machine learning: A survey.
\newblock \emph{ACM Computing Surveys (CSUR)}, 54\penalty0 (11s):\penalty0 1--37, 2022.

\bibitem[Huang et~al.(2017)Huang, Liu, Van Der~Maaten, and Weinberger]{huang2017densely}
Gao Huang, Zhuang Liu, Laurens Van Der~Maaten, and Kilian~Q Weinberger.
\newblock Densely connected convolutional networks.
\newblock In \emph{Proceedings of the IEEE conference on computer vision and pattern recognition}, pages 4700--4708, 2017.

\bibitem[Huber et~al.(2021)Huber, Weitz, Andr{\'e}, and Amir]{huber2021local}
Tobias Huber, Katharina Weitz, Elisabeth Andr{\'e}, and Ofra Amir.
\newblock Local and global explanations of agent behavior: Integrating strategy summaries with saliency maps.
\newblock \emph{Artificial Intelligence}, 301:\penalty0 103571, 2021.

\bibitem[Ifty et~al.(2024)Ifty, Shafin, Shahriar, and Towhid]{ifty2024explainable}
Tanzina~Taher Ifty, Saleh~Ahmed Shafin, Shoeb~Mohammad Shahriar, and Tashfia Towhid.
\newblock Explainable lung disease classification from chest x-ray images utilizing deep learning and xai.
\newblock \emph{arXiv preprint arXiv:2404.11428}, 2024.

\bibitem[Jacovi and Goldberg(2020)]{jacovi2020towards}
Alon Jacovi and Yoav Goldberg.
\newblock Towards faithfully interpretable nlp systems: How should we define and evaluate faithfulness?
\newblock \emph{arXiv preprint arXiv:2004.03685}, 2020.

\bibitem[Jacovi and Goldberg(2021)]{jacovi2021aligning}
Alon Jacovi and Yoav Goldberg.
\newblock Aligning faithful interpretations with their social attribution.
\newblock \emph{Transactions of the Association for Computational Linguistics}, 9:\penalty0 294--310, 2021.

\bibitem[Janizek et~al.(2021)Janizek, Sturmfels, and Lee]{janizek2021explaining}
Joseph~D Janizek, Pascal Sturmfels, and Su-In Lee.
\newblock Explaining explanations: Axiomatic feature interactions for deep networks.
\newblock \emph{Journal of Machine Learning Research}, 22\penalty0 (104):\penalty0 1--54, 2021.

\bibitem[Jentzen et~al.(2023)Jentzen, Kuckuck, and von Wurstemberger]{jentzen2023mathematicalintroductiondeeplearning}
Arnulf Jentzen, Benno Kuckuck, and Philippe von Wurstemberger.
\newblock Mathematical introduction to deep learning: Methods, implementations, and theory, 2023.
\newblock URL \url{https://arxiv.org/abs/2310.20360}.

\bibitem[Ju et~al.(2021)Ju, Zhang, Yang, Jiang, Liu, and Zhao]{ju2021logic}
Yiming Ju, Yuanzhe Zhang, Zhao Yang, Zhongtao Jiang, Kang Liu, and Jun Zhao.
\newblock Logic traps in evaluating attribution scores.
\newblock \emph{arXiv preprint arXiv:2109.05463}, 2021.

\bibitem[Khalid et~al.(2023)Khalid, Qayyum, Bilal, Al-Fuqaha, and Qadir]{khalid2023privacy}
Nazish Khalid, Adnan Qayyum, Muhammad Bilal, Ala Al-Fuqaha, and Junaid Qadir.
\newblock Privacy-preserving artificial intelligence in healthcare: Techniques and applications.
\newblock \emph{Computers in Biology and Medicine}, 158:\penalty0 106848, 2023.

\bibitem[Klabunde et~al.(2024)Klabunde, Schumacher, Strohmaier, and Lemmerich]{klabunde2024similarityneuralnetworkmodels}
Max Klabunde, Tobias Schumacher, Markus Strohmaier, and Florian Lemmerich.
\newblock Similarity of neural network models: A survey of functional and representational measures, 2024.
\newblock URL \url{https://arxiv.org/abs/2305.06329}.

\bibitem[Kokhlikyan et~al.(2021)Kokhlikyan, Miglani, Alsallakh, Martin, and Reblitz-Richardson]{kokhlikyan2021investigating}
Narine Kokhlikyan, Vivek Miglani, Bilal Alsallakh, Miguel Martin, and Orion Reblitz-Richardson.
\newblock Investigating sanity checks for saliency maps with image and text classification.
\newblock \emph{arXiv preprint arXiv:2106.07475}, 2021.

\bibitem[Kornblith et~al.(2019)Kornblith, Norouzi, Lee, and Hinton]{kornblith2019similarity}
Simon Kornblith, Mohammad Norouzi, Honglak Lee, and Geoffrey Hinton.
\newblock Similarity of neural network representations revisited.
\newblock In \emph{International conference on machine learning}, pages 3519--3529. PMLR, 2019.

\bibitem[Krishna et~al.(2022)Krishna, Han, Gu, Pombra, Jabbari, Wu, and Lakkaraju]{krishna2022disagreementproblemexplainablemachine}
Satyapriya Krishna, Tessa Han, Alex Gu, Javin Pombra, Shahin Jabbari, Steven Wu, and Himabindu Lakkaraju.
\newblock The disagreement problem in explainable machine learning: A practitioner's perspective, 2022.
\newblock URL \url{https://arxiv.org/abs/2202.01602}.

\bibitem[Kumar et~al.(2020)Kumar, Venkatasubramanian, Scheidegger, and Friedler]{kumar2020problems}
I~Elizabeth Kumar, Suresh Venkatasubramanian, Carlos Scheidegger, and Sorelle Friedler.
\newblock Problems with shapley-value-based explanations as feature importance measures.
\newblock In \emph{International conference on machine learning}, pages 5491--5500. PMLR, 2020.

\bibitem[Li et~al.(2021)Li, Zhang, Zhou, Fu, Xia, and Hu]{li2021deep}
Junbing Li, Changqing Zhang, Joey~Tianyi Zhou, Huazhu Fu, Shuyin Xia, and Qinghua Hu.
\newblock Deep-lift: Deep label-specific feature learning for image annotation.
\newblock \emph{IEEE transactions on Cybernetics}, 52\penalty0 (8):\penalty0 7732--7741, 2021.

\bibitem[Li et~al.(2023)Li, Du, Chen, Chai, Lakkaraju, and Xiong]{li2023mathcal}
Xuhong Li, Mengnan Du, Jiamin Chen, Yekun Chai, Himabindu Lakkaraju, and Haoyi Xiong.
\newblock M4: A unified xai benchmark for faithfulness evaluation of feature attribution methods across metrics, modalities and models.
\newblock \emph{Advances in Neural Information Processing Systems}, 36:\penalty0 1630--1643, 2023.

\bibitem[Lowy et~al.(2024)Lowy, Li, Liu, Koike-Akino, Parsons, and Wang]{lowy2024does}
Andrew Lowy, Zhuohang Li, Jing Liu, Toshiaki Koike-Akino, Kieran Parsons, and Ye~Wang.
\newblock Why does differential privacy with large epsilon defend against practical membership inference attacks?
\newblock \emph{arXiv preprint arXiv:2402.09540}, 2024.

\bibitem[Lucieri et~al.(2023)Lucieri, Dengel, and Ahmed]{lucieri2023translating}
Adriano Lucieri, Andreas Dengel, and Sheraz Ahmed.
\newblock Translating theory into practice: assessing the privacy implications of concept-based explanations for biomedical ai.
\newblock \emph{Frontiers in Bioinformatics}, 3:\penalty0 1194993, 2023.

\bibitem[Lundberg and Lee(2017)]{lundberg2017unified}
Scott~M Lundberg and Su-In Lee.
\newblock A unified approach to interpreting model predictions.
\newblock \emph{Advances in neural information processing systems}, 30, 2017.

\bibitem[Lundberg et~al.(2020)Lundberg, Erion, Chen, DeGrave, Prutkin, Nair, Katz, Himmelfarb, Bansal, and Lee]{lundberg2020local}
Scott~M Lundberg, Gabriel Erion, Hugh Chen, Alex DeGrave, Jordan~M Prutkin, Bala Nair, Ronit Katz, Jonathan Himmelfarb, Nisha Bansal, and Su-In Lee.
\newblock From local explanations to global understanding with explainable ai for trees.
\newblock \emph{Nature machine intelligence}, 2\penalty0 (1):\penalty0 56--67, 2020.

\bibitem[Lyu et~al.(2024)Lyu, Apidianaki, and Callison-Burch]{lyu2024towards}
Qing Lyu, Marianna Apidianaki, and Chris Callison-Burch.
\newblock Towards faithful model explanation in nlp: A survey.
\newblock \emph{Computational Linguistics}, pages 1--67, 2024.

\bibitem[Mochaourab et~al.(2021)Mochaourab, Sinha, Greenstein, and Papapetrou]{mochaourab2021robust}
Rami Mochaourab, Sugandh Sinha, Stanley Greenstein, and Panagiotis Papapetrou.
\newblock Robust counterfactual explanations for privacy-preserving svm.
\newblock In \emph{International Conference on Machine Learning (ICML 2021), Workshop on Socially Responsible Machine Learning}, 2021.

\bibitem[Mohammadi and Petridis(2022)]{MOHAMMADI2022103515}
Ali Mohammadi and Giorgis Petridis.
\newblock Almost orthogonal subsets of vector spaces over finite fields.
\newblock \emph{European Journal of Combinatorics}, 103:\penalty0 103515, 2022.
\newblock ISSN 0195-6698.
\newblock \doi{https://doi.org/10.1016/j.ejc.2022.103515}.
\newblock URL \url{https://www.sciencedirect.com/science/article/pii/S0195669822000117}.

\bibitem[Montavon et~al.(2019)Montavon, Binder, Lapuschkin, Samek, and M{\"u}ller]{montavon2019layer}
Gr{\'e}goire Montavon, Alexander Binder, Sebastian Lapuschkin, Wojciech Samek, and Klaus-Robert M{\"u}ller.
\newblock Layer-wise relevance propagation: an overview.
\newblock \emph{Explainable AI: interpreting, explaining and visualizing deep learning}, pages 193--209, 2019.

\bibitem[Naresh et~al.(2023)Naresh, Thamarai, and Allavarpu]{naresh2023privacy}
Vankamamidi~S Naresh, Muthusamy Thamarai, and VVL~Divakar Allavarpu.
\newblock Privacy-preserving deep learning in medical informatics: applications, challenges, and solutions.
\newblock \emph{Artificial Intelligence Review}, 56\penalty0 (Suppl 1):\penalty0 1199--1241, 2023.

\bibitem[{National Institute of Allergy and Infectious Diseases}()]{NIAID_TB_Portal}
{National Institute of Allergy and Infectious Diseases}.
\newblock Niaid tb portal program dataset.
\newblock URL \url{https://tbportals.niaid.nih.gov/}.

\bibitem[Neloy and Turgeon(2024)]{neloy2024comprehensive}
Asif~Ahmed Neloy and Maxime Turgeon.
\newblock A comprehensive study of auto-encoders for anomaly detection: Efficiency and trade-offs.
\newblock \emph{Machine Learning with Applications}, page 100572, 2024.

\bibitem[Nguyen and Mart{\'\i}nez(2020)]{nguyen2020quantitative}
An-phi Nguyen and Mar{\'\i}a~Rodr{\'\i}guez Mart{\'\i}nez.
\newblock On quantitative aspects of model interpretability.
\newblock \emph{arXiv preprint arXiv:2007.07584}, 2020.

\bibitem[Nguyen et~al.(2022)Nguyen, Lam, Le, Pham, Tran, Nguyen, Le, Pham, Tong, Dinh, et~al.]{nguyen2022vindr}
Ha~Q Nguyen, Khanh Lam, Linh~T Le, Hieu~H Pham, Dat~Q Tran, Dung~B Nguyen, Dung~D Le, Chi~M Pham, Hang~TT Tong, Diep~H Dinh, et~al.
\newblock Vindr-cxr: An open dataset of chest x-rays with radiologist’s annotations.
\newblock \emph{Scientific Data}, 9\penalty0 (1):\penalty0 429, 2022.

\bibitem[Nguyen et~al.(2024)Nguyen, Huynh, Ren, Nguyen, Nguyen, Yin, and Nguyen]{nguyen2024surveyprivacypreservingmodelexplanations}
Thanh~Tam Nguyen, Thanh~Trung Huynh, Zhao Ren, Thanh~Toan Nguyen, Phi~Le Nguyen, Hongzhi Yin, and Quoc Viet~Hung Nguyen.
\newblock A survey of privacy-preserving model explanations: Privacy risks, attacks, and countermeasures, 2024.
\newblock URL \url{https://arxiv.org/abs/2404.00673}.

\bibitem[Oliynyk et~al.(2023)Oliynyk, Mayer, and Rauber]{oliynyk2023know}
Daryna Oliynyk, Rudolf Mayer, and Andreas Rauber.
\newblock I know what you trained last summer: A survey on stealing machine learning models and defences.
\newblock \emph{ACM Computing Surveys}, 55\penalty0 (14s):\penalty0 1--41, 2023.

\bibitem[Papernot et~al.(2018)Papernot, Song, Mironov, Raghunathan, Talwar, and Úlfar Erlingsson]{papernot2018scalableprivatelearningpate}
Nicolas Papernot, Shuang Song, Ilya Mironov, Ananth Raghunathan, Kunal Talwar, and Úlfar Erlingsson.
\newblock Scalable private learning with pate, 2018.
\newblock URL \url{https://arxiv.org/abs/1802.08908}.

\bibitem[Patel et~al.(2022)Patel, Shokri, and Zick]{patel2022model}
Neel Patel, Reza Shokri, and Yair Zick.
\newblock Model explanations with differential privacy.
\newblock In \emph{Proceedings of the 2022 ACM Conference on Fairness, Accountability, and Transparency}, pages 1895--1904, 2022.

\bibitem[Patel(2020)]{prashant2682020chestxray}
Prashant Patel.
\newblock Chest x-ray (covid-19 \& pneumonia), 2020.
\newblock URL \url{https://www.kaggle.com/datasets/prashant268/chest-xray-covid19-pneumonia}.

\bibitem[Petsiuk et~al.(2018)Petsiuk, Das, and Saenko]{petsiuk2018rise}
Vitali Petsiuk, Abir Das, and Kate Saenko.
\newblock Rise: Randomized input sampling for explanation of black-box models.
\newblock \emph{arXiv preprint arXiv:1806.07421}, 2018.

\bibitem[Ponomareva et~al.(2023)Ponomareva, Hazimeh, Kurakin, Xu, Denison, McMahan, Vassilvitskii, Chien, and Thakurta]{Ponomareva_2023}
Natalia Ponomareva, Hussein Hazimeh, Alex Kurakin, Zheng Xu, Carson Denison, H.~Brendan McMahan, Sergei Vassilvitskii, Steve Chien, and Abhradeep~Guha Thakurta.
\newblock How to dp-fy ml: A practical guide to machine learning with differential privacy.
\newblock \emph{Journal of Artificial Intelligence Research}, 77:\penalty0 1113–1201, July 2023.
\newblock ISSN 1076-9757.
\newblock \doi{10.1613/jair.1.14649}.
\newblock URL \url{http://dx.doi.org/10.1613/jair.1.14649}.

\bibitem[Puccetti(2022)]{puccetti2022measuring}
Giovanni Puccetti.
\newblock Measuring linear correlation between random vectors.
\newblock \emph{Information Sciences}, 607:\penalty0 1328--1347, 2022.

\bibitem[Pulido-Gaytan et~al.(2020)Pulido-Gaytan, Tchernykh, Cort{\'e}s-Mendoza, Babenko, and Radchenko]{pulido2020survey}
Luis~Bernardo Pulido-Gaytan, Andrei Tchernykh, Jorge~M Cort{\'e}s-Mendoza, Mikhail Babenko, and Gleb Radchenko.
\newblock A survey on privacy-preserving machine learning with fully homomorphic encryption.
\newblock In \emph{Latin American High Performance Computing Conference}, pages 115--129. Springer, 2020.

\bibitem[Rahman et~al.(2020)Rahman, Khandakar, Kadir, Islam, Islam, Mazhar, Hamid, Islam, Kashem, Mahbub, et~al.]{rahman2020reliable}
Tawsifur Rahman, Amith Khandakar, Muhammad~Abdul Kadir, Khandaker~Rejaul Islam, Khandakar~F Islam, Rashid Mazhar, Tahir Hamid, Mohammad~Tariqul Islam, Saad Kashem, Zaid~Bin Mahbub, et~al.
\newblock Reliable tuberculosis detection using chest x-ray with deep learning, segmentation and visualization.
\newblock \emph{Ieee Access}, 8:\penalty0 191586--191601, 2020.

\bibitem[Ribeiro et~al.(2016)Ribeiro, Singh, and Guestrin]{ribeiro2016should}
Marco~Tulio Ribeiro, Sameer Singh, and Carlos Guestrin.
\newblock " why should i trust you?" explaining the predictions of any classifier.
\newblock In \emph{Proceedings of the 22nd ACM SIGKDD international conference on knowledge discovery and data mining}, pages 1135--1144, 2016.

\bibitem[Rieger and Hansen(2020)]{rieger2020irof}
Laura Rieger and Lars~Kai Hansen.
\newblock Irof: a low resource evaluation metric for explanation methods.
\newblock \emph{arXiv preprint arXiv:2003.08747}, 2020.

\bibitem[Rigaki and Garcia(2023)]{Rigaki_2023}
Maria Rigaki and Sebastian Garcia.
\newblock A survey of privacy attacks in machine learning.
\newblock \emph{ACM Computing Surveys}, 56\penalty0 (4):\penalty0 1–34, November 2023.
\newblock ISSN 1557-7341.
\newblock \doi{10.1145/3624010}.
\newblock URL \url{http://dx.doi.org/10.1145/3624010}.

\bibitem[Saeed and Omlin(2023)]{saeed2023explainable}
Waddah Saeed and Christian Omlin.
\newblock Explainable ai (xai): A systematic meta-survey of current challenges and future opportunities.
\newblock \emph{Knowledge-Based Systems}, 263:\penalty0 110273, 2023.

\bibitem[Saifullah et~al.(2024)Saifullah, Mercier, Lucieri, Dengel, and Ahmed]{saifullah2024privacy}
Saifullah Saifullah, Dominique Mercier, Adriano Lucieri, Andreas Dengel, and Sheraz Ahmed.
\newblock The privacy-explainability trade-off: unraveling the impacts of differential privacy and federated learning on attribution methods.
\newblock \emph{Frontiers in Artificial Intelligence}, 7:\penalty0 1236947, 2024.

\bibitem[Samek et~al.(2016)Samek, Binder, Montavon, Lapuschkin, and M{\"u}ller]{samek2016evaluating}
Wojciech Samek, Alexander Binder, Gr{\'e}goire Montavon, Sebastian Lapuschkin, and Klaus-Robert M{\"u}ller.
\newblock Evaluating the visualization of what a deep neural network has learned.
\newblock \emph{IEEE transactions on neural networks and learning systems}, 28\penalty0 (11):\penalty0 2660--2673, 2016.

\bibitem[Samek et~al.(2017)Samek, Wiegand, and Müller]{samek2017explainableartificialintelligenceunderstanding}
Wojciech Samek, Thomas Wiegand, and Klaus-Robert Müller.
\newblock Explainable artificial intelligence: Understanding, visualizing and interpreting deep learning models, 2017.
\newblock URL \url{https://arxiv.org/abs/1708.08296}.

\bibitem[Saxena et~al.(2022)Saxena, Singh, Tiwary, Mittal, and Jain]{saxena2022artificial}
Pranshu Saxena, Sanjay~Kumar Singh, Gyanendra Tiwary, Yush Mittal, and Ishika Jain.
\newblock An artificial intelligence technique for covid-19 detection with explainability using lungs x-ray images.
\newblock In \emph{2022 IEEE International Conference on Distributed Computing and Electrical Circuits and Electronics (ICDCECE)}, pages 1--6. IEEE, 2022.

\bibitem[Selvaraju et~al.(2019)Selvaraju, Cogswell, Das, Vedantam, Parikh, and Batra]{Selvaraju_2019}
Ramprasaath~R. Selvaraju, Michael Cogswell, Abhishek Das, Ramakrishna Vedantam, Devi Parikh, and Dhruv Batra.
\newblock Grad-cam: Visual explanations from deep networks via gradient-based localization.
\newblock \emph{International Journal of Computer Vision}, 128\penalty0 (2):\penalty0 336–359, October 2019.
\newblock ISSN 1573-1405.
\newblock \doi{10.1007/s11263-019-01228-7}.
\newblock URL \url{http://dx.doi.org/10.1007/s11263-019-01228-7}.

\bibitem[Shokri et~al.(2017)Shokri, Stronati, Song, and Shmatikov]{shokri2017membership}
Reza Shokri, Marco Stronati, Congzheng Song, and Vitaly Shmatikov.
\newblock Membership inference attacks against machine learning models.
\newblock In \emph{2017 IEEE symposium on security and privacy (SP)}, pages 3--18. IEEE, 2017.

\bibitem[Shokri et~al.(2021)Shokri, Strobel, and Zick]{shokri2021privacy}
Reza Shokri, Martin Strobel, and Yair Zick.
\newblock On the privacy risks of model explanations.
\newblock In \emph{Proceedings of the 2021 AAAI/ACM Conference on AI, Ethics, and Society}, pages 231--241, 2021.

\bibitem[Smilkov et~al.(2017)Smilkov, Thorat, Kim, Vi{\'e}gas, and Wattenberg]{smilkov2017smoothgrad}
Daniel Smilkov, Nikhil Thorat, Been Kim, Fernanda Vi{\'e}gas, and Martin Wattenberg.
\newblock Smoothgrad: removing noise by adding noise.
\newblock \emph{arXiv preprint arXiv:1706.03825}, 2017.

\bibitem[Subramani et~al.(2021)Subramani, Vadivelu, and Kamath]{subramani2021enabling}
Pranav Subramani, Nicholas Vadivelu, and Gautam Kamath.
\newblock Enabling fast differentially private sgd via just-in-time compilation and vectorization.
\newblock \emph{Advances in Neural Information Processing Systems}, 34:\penalty0 26409--26421, 2021.

\bibitem[Sundararajan and Najmi(2020)]{sundararajan2020many}
Mukund Sundararajan and Amir Najmi.
\newblock The many shapley values for model explanation.
\newblock In \emph{International conference on machine learning}, pages 9269--9278. PMLR, 2020.

\bibitem[Sundararajan et~al.(2017)Sundararajan, Taly, and Yan]{sundararajan2017axiomatic}
Mukund Sundararajan, Ankur Taly, and Qiqi Yan.
\newblock Axiomatic attribution for deep networks.
\newblock In \emph{International conference on machine learning}, pages 3319--3328. PMLR, 2017.

\bibitem[Suriyakumar et~al.(2021)Suriyakumar, Papernot, Goldenberg, and Ghassemi]{suriyakumar2021chasing}
Vinith~M Suriyakumar, Nicolas Papernot, Anna Goldenberg, and Marzyeh Ghassemi.
\newblock Chasing your long tails: Differentially private prediction in health care settings.
\newblock In \emph{Proceedings of the 2021 ACM Conference on Fairness, Accountability, and Transparency}, pages 723--734, 2021.

\bibitem[Sweeney(2015)]{sweeney2015only}
Latanya Sweeney.
\newblock Only you, your doctor, and many others may know.
\newblock \emph{Technology Science}, 2015092903\penalty0 (9):\penalty0 29, 2015.

\bibitem[Talaei~Khoei et~al.(2023)Talaei~Khoei, Ould~Slimane, and Kaabouch]{talaei2023deep}
Tala Talaei~Khoei, Hadjar Ould~Slimane, and Naima Kaabouch.
\newblock Deep learning: Systematic review, models, challenges, and research directions.
\newblock \emph{Neural Computing and Applications}, 35\penalty0 (31):\penalty0 23103--23124, 2023.

\bibitem[Tan and Le(2021)]{tan2021efficientnetv2}
Mingxing Tan and Quoc Le.
\newblock Efficientnetv2: Smaller models and faster training.
\newblock In \emph{International conference on machine learning}, pages 10096--10106. PMLR, 2021.

\bibitem[Theiner et~al.(2022)Theiner, M{\"u}ller-Budack, and Ewerth]{theiner2022interpretable}
Jonas Theiner, Eric M{\"u}ller-Budack, and Ralph Ewerth.
\newblock Interpretable semantic photo geolocation.
\newblock In \emph{Proceedings of the IEEE/CVF Winter Conference on Applications of Computer Vision}, pages 750--760, 2022.

\bibitem[Thomson(1975)]{thomson1975right}
Judith~Jarvis Thomson.
\newblock The right to privacy.
\newblock \emph{Philosophy \& Public Affairs}, pages 295--314, 1975.

\bibitem[Tursynbek et~al.(2020)Tursynbek, Petiushko, and Oseledets]{tursynbek2020robustness}
Nurislam Tursynbek, Aleksandr Petiushko, and Ivan Oseledets.
\newblock Robustness threats of differential privacy.
\newblock \emph{arXiv preprint arXiv:2012.07828}, 2020.

\bibitem[Veale et~al.(2018)Veale, Binns, and Edwards]{veale2018algorithms}
Michael Veale, Reuben Binns, and Lilian Edwards.
\newblock Algorithms that remember: model inversion attacks and data protection law.
\newblock \emph{Philosophical Transactions of the Royal Society A: Mathematical, Physical and Engineering Sciences}, 376\penalty0 (2133):\penalty0 20180083, 2018.

\bibitem[Vredenburgh(2022)]{vredenburgh2022right}
Kate Vredenburgh.
\newblock The right to explanation.
\newblock \emph{Journal of Political Philosophy}, 30\penalty0 (2):\penalty0 209--229, 2022.

\bibitem[Wang et~al.(2021)Wang, Fu, Li, Khisti, Zemel, and Makhzani]{wang2021variational}
Kuan-Chieh Wang, Yan Fu, Ke~Li, Ashish Khisti, Richard Zemel, and Alireza Makhzani.
\newblock Variational model inversion attacks.
\newblock \emph{Advances in Neural Information Processing Systems}, 34:\penalty0 9706--9719, 2021.

\bibitem[Wang et~al.(2004)Wang, Bovik, Sheikh, and Simoncelli]{wang2004image}
Zhou Wang, Alan~C Bovik, Hamid~R Sheikh, and Eero~P Simoncelli.
\newblock Image quality assessment: from error visibility to structural similarity.
\newblock \emph{IEEE transactions on image processing}, 13\penalty0 (4):\penalty0 600--612, 2004.

\bibitem[Wyner(1967)]{6771362}
A.~D. Wyner.
\newblock Random packings and coverings of the unit n-sphere.
\newblock \emph{The Bell System Technical Journal}, 46\penalty0 (9):\penalty0 2111--2118, 1967.
\newblock \doi{10.1002/j.1538-7305.1967.tb04246.x}.

\bibitem[Yang et~al.(2022)Yang, Feng, Zhou, Chen, and Hu]{yang2022differentially}
Fan Yang, Qizhang Feng, Kaixiong Zhou, Jiahao Chen, and Xia Hu.
\newblock Differentially private counterfactuals via functional mechanism.
\newblock \emph{arXiv preprint arXiv:2208.02878}, 2022.

\bibitem[Yeh et~al.(2019)Yeh, Hsieh, Suggala, Inouye, and Ravikumar]{yeh2019fidelity}
Chih-Kuan Yeh, Cheng-Yu Hsieh, Arun Suggala, David~I Inouye, and Pradeep~K Ravikumar.
\newblock On the (in) fidelity and sensitivity of explanations.
\newblock \emph{Advances in neural information processing systems}, 32, 2019.

\bibitem[Yona and Greenfeld(2021)]{yona2021revisiting}
Gal Yona and Daniel Greenfeld.
\newblock Revisiting sanity checks for saliency maps.
\newblock \emph{arXiv preprint arXiv:2110.14297}, 2021.

\bibitem[Yousefpour et~al.(2022)Yousefpour, Shilov, Sablayrolles, Testuggine, Prasad, Malek, Nguyen, Ghosh, Bharadwaj, Zhao, Cormode, and Mironov]{yousefpour2022opacususerfriendlydifferentialprivacy}
Ashkan Yousefpour, Igor Shilov, Alexandre Sablayrolles, Davide Testuggine, Karthik Prasad, Mani Malek, John Nguyen, Sayan Ghosh, Akash Bharadwaj, Jessica Zhao, Graham Cormode, and Ilya Mironov.
\newblock Opacus: User-friendly differential privacy library in pytorch, 2022.
\newblock URL \url{https://arxiv.org/abs/2109.12298}.

\bibitem[Zhang et~al.(2018)Zhang, Bargal, Lin, Brandt, Shen, and Sclaroff]{zhang2018top}
Jianming Zhang, Sarah~Adel Bargal, Zhe Lin, Jonathan Brandt, Xiaohui Shen, and Stan Sclaroff.
\newblock Top-down neural attention by excitation backprop.
\newblock \emph{International Journal of Computer Vision}, 126\penalty0 (10):\penalty0 1084--1102, 2018.

\bibitem[Zheng et~al.(2024)Zheng, Zhang, Zhang, Song, Zhou, and Han]{zheng2024rethinking}
Yu~Zheng, Wenchao Zhang, Yonggang Zhang, Wei Song, Kai Zhou, and Bo~Han.
\newblock Rethinking improved privacy-utility trade-off with pre-existing knowledge for dp training.
\newblock \emph{arXiv preprint arXiv:2409.03344}, 2024.

\bibitem[Zhou et~al.(2021)Zhou, Gandomi, Chen, and Holzinger]{zhou2021evaluating}
Jianlong Zhou, Amir~H Gandomi, Fang Chen, and Andreas Holzinger.
\newblock Evaluating the quality of machine learning explanations: A survey on methods and metrics.
\newblock \emph{Electronics}, 10\penalty0 (5):\penalty0 593, 2021.

\bibitem[Zhu et~al.(2024)Zhu, Ma, Cheng, Zhang, Zhang, and Liu]{zhu2024openworldmachinelearningreview}
Fei Zhu, Shijie Ma, Zhen Cheng, Xu-Yao Zhang, Zhaoxiang Zhang, and Cheng-Lin Liu.
\newblock Open-world machine learning: A review and new outlooks, 2024.
\newblock URL \url{https://arxiv.org/abs/2403.01759}.

\end{thebibliography}

\newpage

\section*{Appendix A.}\label{App A}

\begin{figure}[h]
    \subfloat[DenseNet, Saliency]{%
        \includegraphics[width=0.48\textwidth]{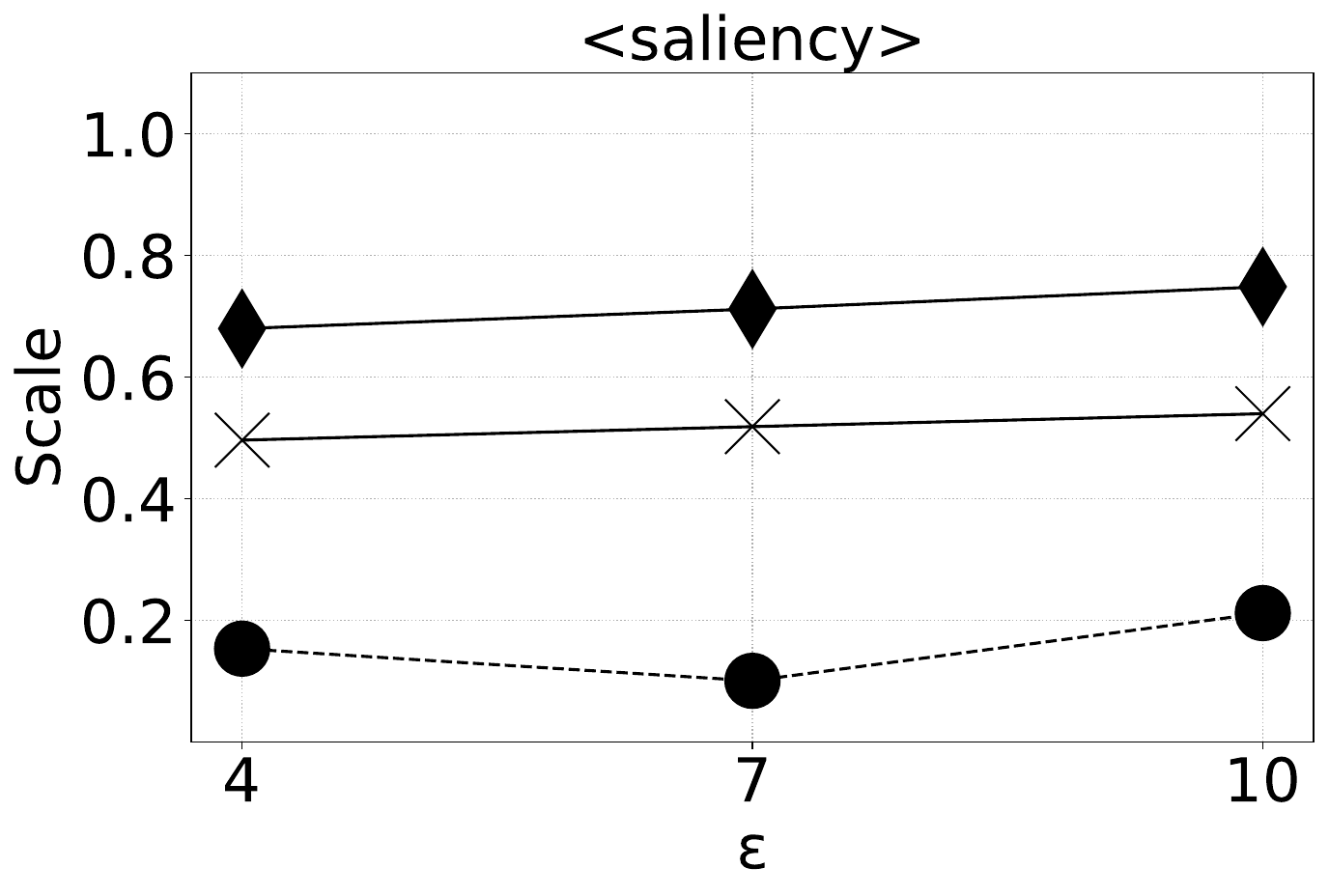}
    }\hfill
    \subfloat[ResNet, Saliency]{%
        \includegraphics[width=0.48\textwidth]{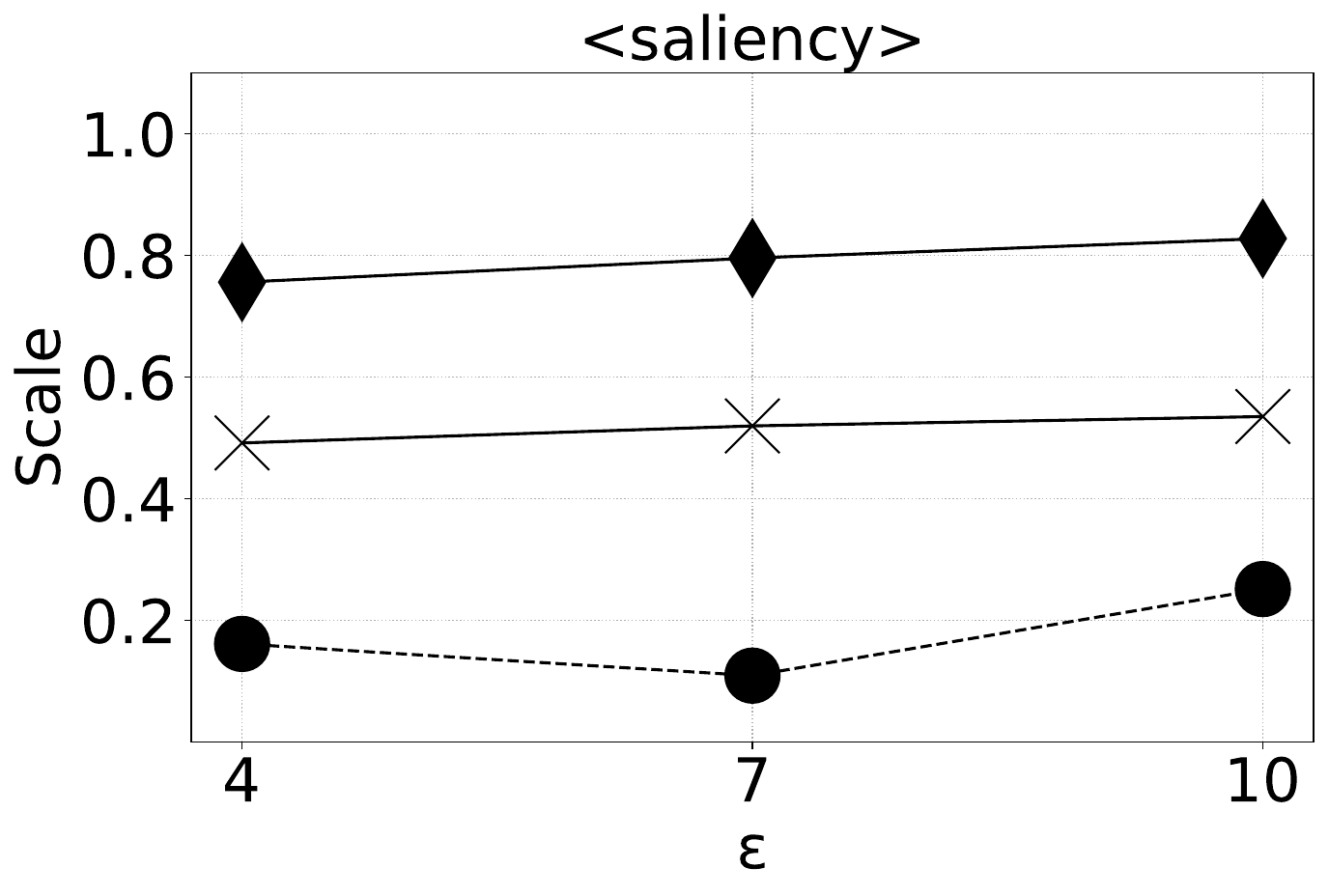}
    }

    \vspace{0.5cm} % Add vertical space between rows if needed

    \subfloat[DenseNet, SmoothGrad]{%
        \includegraphics[width=0.48\textwidth]{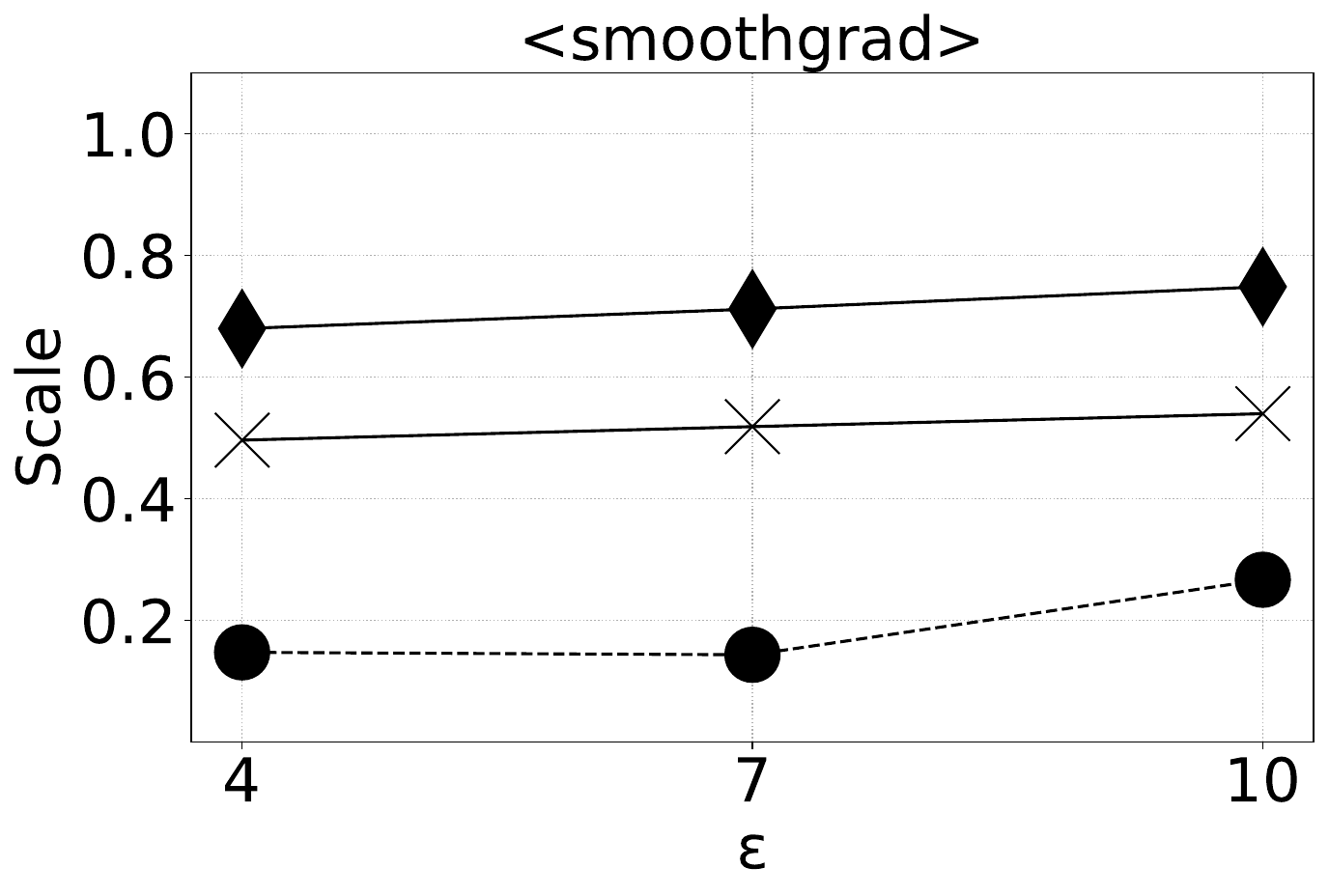}
    }\hfill
    \subfloat[ResNet, SmoothGrad]{%
        \includegraphics[width=0.48\textwidth]{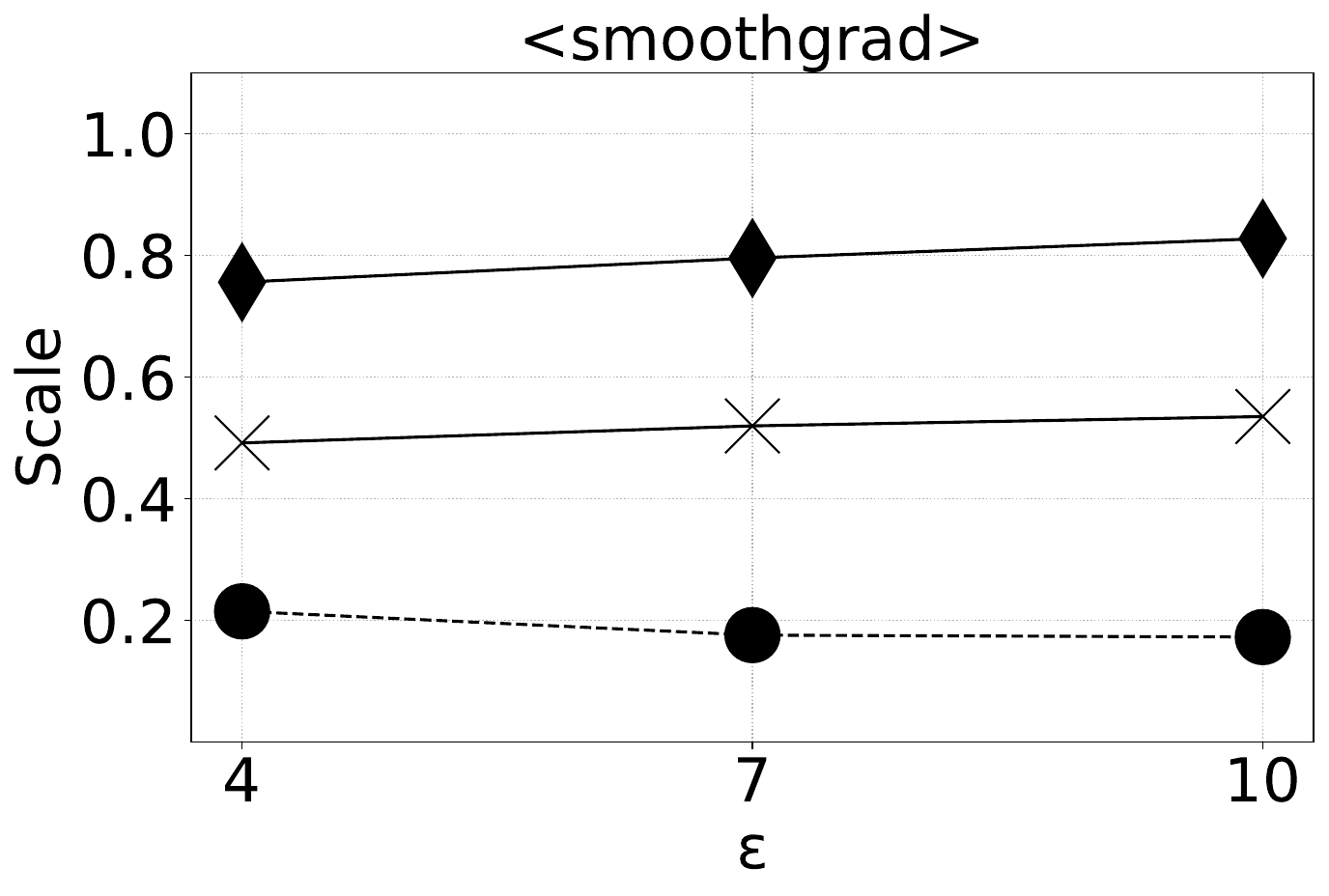}
    }

    \caption{Analysis on CIFAR-10: -\raisebox{-0.5ex}{\scalebox{2}{•}}-
 \hspace{.05cm} for $PIS_{Avg}$, $-\blacklozenge-$ for $Acc_{\mathcal{M'}/\mathcal{M}}$, and $-\times-$ for $Agreement$ between non-private and private model pair.}
    \label{dp-CIFER}
\end{figure}

We trained ResNet-34 and DenseNet-121 models on the CIFAR-10 dataset using three \(\epsilon\) values (\(4, 7, 10\)), as lower \(\epsilon\) values resulted in harsh privacy-utility trade-off. We weren't able to train EfficientNet-v2 due to its exuberant computational requirements. We evaluated the models on the whole test set. We train all models (both DP and non-DP counterparts) with the exact set of hyperparameters over $50$ epochs. For brevity, we report $Acc_{\mathcal{M}}$ (Table \ref{base-model-cifar}), $Acc_{\mathcal{M'}/\mathcal{M}}$, and $-\times-$ taking all classes together for \textit{Perf Comp.} and \textit{Agreement} in Figure \ref{dp-CIFER}.

However, in this case as well \texttt{Integrated Gradients} and \texttt{Grad-Shap} yield $30\%$ as mean DS score, and the rest of the explainers also do not obtain $PIS_{Avg}>0.3$ (Figure~\ref{dp-CIFER}). Furthermore, here also we do not find any (apparent) $PIS_{Avg}$ follows with \textit{Perf Comp.} and/or \textit{Agreement}. All in all, the results from our primary experiment are sufficiently comparable here as well. We will release the weights of these models upon publication.

\begin{figure}[ht]
    \centering       
    \includegraphics[width=\textwidth]{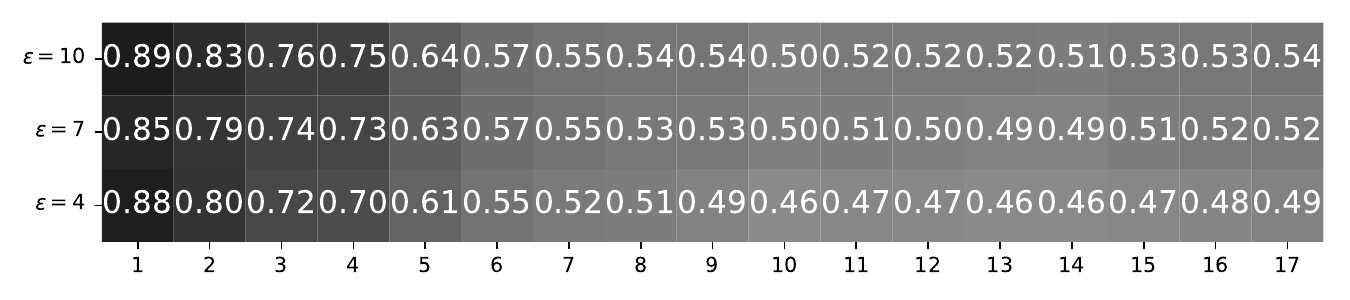}
    \caption{dCKA heatmap for ResNet-34 for CIFAR-10.}
    \label{Investigating_resnet}
\end{figure}

\begin{figure}[h]
    \centering
    \includegraphics[width=\textwidth]{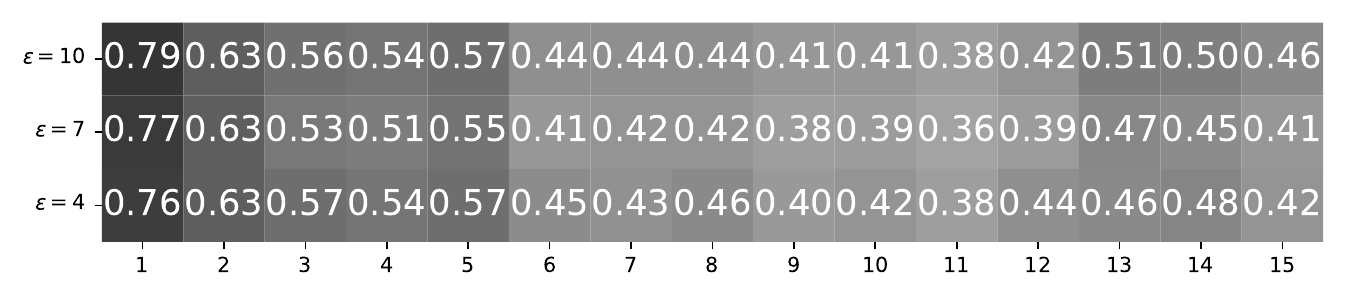}
    \caption{dCKA heatmap for DenseNet-121 for CIFAR-10.}
    \label{Investigating_densenet}
 \end{figure}

For ResNet-34, we observed that sensitivity is independent across all layers for $\epsilon = 10$. However, only the last 2-3 layers for other $\epsilon$ values for ResNet-34 exhibited independent sensitivity. It indicates that independent sensitivity, even in the last few layers, can potentially make the explanations incomparable across \(\epsilon\) values. In contrast, for DenseNet-121, independent sensitivity was consistently observed for all layers across all $\epsilon$. For both types of models, we obtained fewer than 5--7\% of layers where we couldn't reject the null hypothesis for independence of representation. We have reported the dCKA heatmap for DenseNet-121 and ResNet-34 models in Figure \ref{Investigating_densenet}, \ref{Investigating_resnet}, respectively. Notably, the final layer for both ResNet-34 and DesneNet-121 consistently demonstrated independent sensitivity, rendering \texttt{Grad-CAM} unsuitable for generating comparable explanations. Consequently, we focused on other explainability methods in Figure \ref{dp-CIFER}.

\begin{table}[h!]
\centering
\begin{tabular}{c|c}

\textbf{Model} & \textbf{Acc (\%)} \\
\hline
DenseNet-121 & 72.82 \\
ResNet-34 & 66.61 \\

\end{tabular}
\caption{Non-private models' accuracy}
\label{base-model-cifar}
\end{table}

\newpage
\section*{Appendix B.}\label{AppC}

\paragraph{LIME (Local Interpretable Model-Agnostic Explanations):} LIME \citep{ribeiro2016should} approximates a complex model locally using a simpler interpretable model (e.g., linear regression). For an input \(x\), LIME generates perturbed versions of \(x\) and calculates the corresponding predictions to fit a local surrogate model. Mathematically, LIME solves the following optimization problem:
\[
\arg\min_{g \in G} \mathcal{L}(f, g, \pi_x) + \Omega(g),
\]
where \(f\) is the original model, \(g\) is the interpretable surrogate model, \(\pi_x\) is a proximity measure to \(x\), and \(\Omega(g)\) ensures simplicity of \(g\).

\paragraph{SHAP (SHapley Additive exPlanations):} SHAP \citep{lundberg2017unified} explains model predictions based on cooperative game theory. For a prediction \(f(x)\), SHAP attributes contributions to each feature using Shapley values:
\[
\phi_i = \sum_{S \subseteq N \setminus \{i\}} \frac{|S|!(|N| - |S| - 1)!}{|N|!} \left[f(S \cup \{i\}) - f(S)\right],
\]
where \(N\) is the set of all features, \(S\) is a subset of features, and \(f(S)\) is the model's prediction with features in \(S\) included.

\paragraph{Saliency Maps:} Saliency maps visualize the importance of each input feature by computing the gradient of the output \(f(x)\) with respect to the input \(x\):
\[
\text{Saliency}(x_i) = \left|\frac{\partial f(x)}{\partial x_i}\right|.
\]

\paragraph{SmoothGrad:} SmoothGrad \citep{smilkov2017smoothgrad} reduces noise in saliency maps by averaging gradients over multiple noisy samples of the input:
\[
\text{SmoothGrad}(x) = \frac{1}{n} \sum_{i=1}^n \nabla_x f(x + \mathcal{N}(0, \sigma^2)).
\]

\paragraph{Integrated Gradients:} Integrated Gradients \citep{sundararajan2017axiomatic} attribute feature importance by integrating the gradients along the path from a baseline input \(x'\) to the actual input \(x\):
\[
\text{IG}_i(x) = (x_i - x'_i) \int_{\alpha=0}^1 \frac{\partial f(x' + \alpha(x - x'))}{\partial x_i} d\alpha.
\]

\paragraph{Grad-CAM:} Grad-CAM \citep{Selvaraju_2019} generates heatmaps for convolutional neural networks by using gradients of the target output with respect to feature maps of a convolutional layer. For a given feature map \(A_k\), the weights are computed as:
\[
\alpha_k^c = \frac{1}{Z} \sum_{i} \sum_{j} \frac{\partial y^c}{\partial A_k^{ij}},
\]
where \(y^c\) is the output score for class \(c\), and \(Z\) is the spatial dimensions of \(A_k\). The Grad-CAM heatmap is:
\[
\text{Grad-CAM} = \text{ReLU}\left(\sum_k \alpha_k^c A_k\right).
\]
\newpage
\section*{Appendix C.}\label{App D}
\subsection{Additional Details on Experimental Setup}
We train the non-private and private models fixing all the hyperparameters (batch size: 128, lr: 0.001, delta (for DP): 0.001) except for the number of epochs, as private models need more computation to learn due to the heavy regularization DP introduces in the training \citep{Ponomareva_2023}. Following \citep{Ponomareva_2023}, we initialised all our models (both non-private and private counterparts) with publicly available pre-trained weights (ImageNet) for \textit{better} convergence. Furthermore, to make a fair comparison we have fixed the number of all hyperparameters in all the private models with different $\epsilon$. We set the no. of epochs as 50 for all private models; it yielded competitive accuracy across model types. However, within $13-16$ epochs, all non-private models achieved accuracy $>95\%$.

However, while training, we discovered an issue in the vanilla architecture of the aforementioned networks: they utilize batch normalization (\texttt{BatchNorm}). Nevertheless, \texttt{BatchNorm} normalizes a sample based on the statistics of the batch it is in. This means the same sample can get different normalized values depending on the other samples in the batch. For differential privacy, each sample's privacy needs to be independently preserved. Since \texttt{BatchNorm} depends on other samples, it violates this principle and leaks information about other samples. \citep{yousefpour2022opacususerfriendlydifferentialprivacy} advises replacing \texttt{BatchNorm} layers with privacy-friendly options like Group Normalization, Layer Normalization, Instance Normalization, etc. From the engineering perspective, we have to select one such replacement that scales with sufficiently large datasets without hampering the privacy bounds. Based on previous empirical evidence \citep{subramani2021enabling}, we replace the \texttt{BatchNorm} layers with \texttt{GroupNorm} layers in all non-private models along with their private counterparts, as \texttt{GroupNorm} does not alter the base architecture drastically, scales well, and adheres to the privacy principle strictly. We utilized the \texttt{Opacus} library for DP-training (\texttt{\url{https://opacus.ai}}). 

We employ the off-the-shelf, publicly available implementations of the explainers from \texttt{Captum} library. (\texttt{\url{https://captum.ai}}).

\subsection{Implementation Details for Section 7}
For (d)CKA, we utilized the publicly available package: \texttt{Simtorch} (\texttt{\url{https://github.com/ykumards/simtorch}}) with default (hyper)parameter selection. For Statistical testing with HSIC, we utilised the publicly available package: \texttt{PyRKHSstats} (\texttt{\url{https://github.com/Black-Swan-ICL/PyRKHSstats}}) with default (hyper)parameter selection except for the default p-value cutoff of $0.01$. We have used a p-value threshold of $0.05$ throughout our experiments. In all our experiments, we considered the whole test set at once as a single batch for dCKA calculation and statistical testing with HSIC. 

We run all our experiments on an NVIDIA DGX workstation, leveraging 1 Tesla V100 32GB GPU. We wrote all experiments in Python 3.10. Our total computational time for all experiments is roughly 81 hours.

\end{document}